\documentclass[journal,12pt,onecolumn]{IEEEtran}

\usepackage{cite}
\usepackage{graphicx}  
\usepackage{url}     
\usepackage{amsmath} 
\usepackage{amssymb}

\newtheorem{definition}{Definition}
\newtheorem{proposition}{Proposition}

\newtheorem{assumption}{Assumption}
\newtheorem{corollary}{Corollary}
\usepackage{color}
\usepackage{bm}
\usepackage[dvipsnames]{xcolor}
\usepackage{booktabs}
\usepackage{multirow}
\usepackage{tabularx}
\usepackage{epstopdf}

\newcommand{\Disc}{\textnormal{Disc}}
\newcommand{\Viab}{\textnormal{Viab}}

\usepackage[linesnumbered, ruled]{algorithm2e}
\SetKwRepeat{Do}{do}{while}%

\usepackage{booktabs}
\newcommand{\ra}[1]{\renewcommand{\arraystretch}{#1}}

\title{\LARGE \bf
Real-Time Control for Autonomous Racing Based on Viability Theory
}

\author{Alexander Liniger and John Lygeros
\thanks{ The authors are with the Automatic Control Laboratory, ETH
Zurich, 8092 Z\"urich, Switzerland; Emails: 
\{liniger, lygeros\}@control.ee.ethz.ch

This manuscript is the preprint of a paper accepted in the IEEE Transaction in Control System Technology and is subject to IEEE copyright. IEEE maintains the sole rights of distribution or publication of the work in all forms and media.}
}

\begin{document}

\maketitle
\thispagestyle{empty}
\pagestyle{empty}

\begin{abstract}
In this paper we consider autonomous driving of miniature race cars. The viability kernel is used to efficiently generate finite look-ahead trajectories that maximize progress while remaining recursively feasible with respect to static obstacles (e.g., stay inside the track). Together with a low-level model predictive controller, this method makes real-time autonomous racing possible. The viability kernel computation is based on space discretization. To make the calculation robust against discretization errors, we propose a novel numerical scheme based on  game theoretical methods, in particular the discriminating kernel. We show that the resulting algorithm provides an inner approximation of the viability kernel and guarantees that, for all states in the cell surrounding a viable grid point, there exists a control that keeps the system within the kernel. The performance of the proposed control method is studied in simulation where we determine the effects of various design choices and parameters and in experiments on an autonomous racing set-up maintained at the Automatic Control Laboratory of ETH Zurich. Both simulation and experimental results suggest that the more conservative approximation using the discriminating kernel results in safer driving style at the cost of a small increase in lap time.
\end{abstract}

\section{Introduction}
Control design for autonomous driving has attracted considerable attention from the research community and has been successfully demonstrated on several occasions, e.g., \cite{Horowitz2000,Buehler2009}. One form of autonomous driving is autonomous racing where the objective is to drive a car as fast as possible around a predefined track \cite{stanford}. A popular control method for autonomous driving is receding horizon optimal control that has successfully been used for both autonomous driving in general, and for autonomous racing in particular \cite{Borrelli2005,Gao2010,DiCairano2010,Rosolia2017,Liniger_2014}. In both cases the requirement for the car to stay on the track is typically encoded through state constraints. Without appropriate modifications, these constraints can lead to a loss of feasibility of the underlying optimization problem when the control algorithm is implemented in receding horizon, which may lead to accidents. In receding horizon control, recursive feasibility can be achieved by imposing terminal set constraints \cite{Mayne2000}. In autonomous driving, however, this issue is often neglected because terminal set constraints can be difficult to compute. In this work, we employ viability theory to derive recursively feasible controllers for autonomous racing.

Viability theory was developed to characterize states of a dynamical system for which solutions exist that satisfy its state constraints \cite{Aubin2009}. Though the use of viability theory in model predictive control has been considered \cite{Grune2011}, most applications of the theory limit themselves to establishing safe/viable regions in the state space  \cite{Kitsios2005a,Seube2000,Moitie2000,Panagou2009,Nilsson2014} and reconstructing safe feedback controls \cite{Tinka2009}. At the heart of all these applications is the computation of the viability kernel, the largest subset of states for which the state constraints can be satisfied indefinitely. The viability kernel is typically approximated numerically using either the viability kernel algorithm introduced in \cite{saintPierre94}, the discriminating kernel algorithm in the case of an additional disturbance input \cite{Cardaliaguet99}, or by exploiting the link to optimal control through viscosity solutions for Hamilton-Jacobi partial differential equations \cite{Lygeros2004,Mitchell2005}. All these numerical methods are based on gridding the state space and hence their computational complexity grows exponentially in the dimension of the state space. To reduce the computational load, the link to reachable set calculation has been exploited in \cite{Maidens2013}, that allows for the development of efficient algorithms  for linear systems \cite{Chutinan1999,Girard2006,Kurzhanskiy2007,Girard2008}. 
Similarly, if the dynamical system is polynomial and the constraints are semi-algebraic sets, gridding can be avoided by using methods based on linear matrix inequality (LMI) hierarchies \cite{Korda2014}. 

In this paper we show how viability theory can be used to guarantee recursive feasibility of a path planner for miniature race cars. Our contributions are threefold: First, we extend the control scheme of \cite{Liniger_2014} to efficiently generate viable trajectories during the path planning stage. Model Predictive Control (MPC) is then used to track the trajectory, generated in the path planning stage which maximizes the progress. We show that apart from guaranteeing safety, resorting to the viability kernel also reduces online computation time, enabling the use of longer prediction horizons. A related approach was proposed in \cite{Mueller2013}, where the authors exhaustively generate trajectories and then discard the infeasible ones a posteriori. Viability theory was used in \cite{Moitie2000,Kalisiak2007} to construct safe trajectories and speed up trajectory planning processes, respectively. Similarly, viability and reachability analysis have also been used in autonomous driving to guarantee safety \cite{Althoff2009,Wood2012,Nilsson2014}. The main difference between those approaches and our method is that we use the viability kernel in a receding horizon fashion to generate viable trajectories only, which not only ensures recursive feasibility but also speeds up the computation.

The second contribution of this paper concerns the computation of the viability kernel. In particular, we propose a method which takes into account the discretization error due to the gridding in the viability kernel algorithm. More specifically, we propose to model the discretization error as an additive uncertainty, and then formulate the viability computation as a dynamic game between the uncertainty and the control input. The victory domain of this game is then computed using the discriminating kernel algorithm \cite{Cardaliaguet99}. This stands in contrast to other algorithms for computing the viability kernel, which mainly establish inner or outer approximation thereof. For example, the effects due to the discretization are considered in \cite{saintPierre94,Cardaliaguet99} by ``extending" the set-valued dynamics, which results in an outer approximation of the viability kernel; an error bound for this approximation is given in \cite{Rieger09}. A different approach was proposed in \cite{Lhommeau2011}, where interval analysis is used to find inner and outer approximations of the viability kernel. Compared to those methods our game theoretical approach, which is an inner approximation of the viability kernel, does not only guarantee viability of the grid points, but also viability of points ``close" to viable grid points. This property is of special interest in real applications, where the state of a system rarely coincides with a grid point.

As a third contribution of this paper, we examine the performance of the proposed controller in an extensive simulation study where we perform a detailed sensitivity analysis with respect to parameters of our viability-based controller. We demonstrate that the proposed controller scheme not only improves the overall performance of the race cars, but also dramatically decreases the online computation time compared to the original controller of \cite{Liniger_2014}; the price to pay for this reduction in online computation is a significant increase in offline computations. Finally we verify the controller's performance experimentally which shows that our viability-based controller is indeed suited for autonomous racing.

We remark that the current paper is an extended version of our previous publication \cite{Liniger_2015}, where we presented partial results of this paper's first contribution.

This paper is organized as follows. Section \ref{sec:HRHC} presents the hierarchical control structure of \cite{Liniger_2014} for the miniature race cars, and  Section \ref{sec:Viab} summarizes the viability and discriminating kernel algorithm of \cite{saintPierre94,Cardaliaguet99}. In Section \ref{sec:spaceDis} the new viability kernel approximation is introduced. Section \ref{sec:Hybrid} formulates the path planning step of the hierarchical controller as a discrete-time system. The viability and discriminating kernel based on this model are then analyzed in Section \ref{sec:viabKernelTrack}. In Sections \ref{sec:sim} and \ref{sec:exp} we study the performance of the resulting controller both in simulation and in experiment. Conclusions and directions of future work are provided in Section \ref{sec:conclusion}. Appendices \ref{app:InnerApprox} and \ref{app:N_m} contain technical results and proofs.

\section{Autonomous racing control hierarchy} \label{sec:HRHC}
\begin{figure}[h]
\centering
\includegraphics[width = 0.65\textwidth]{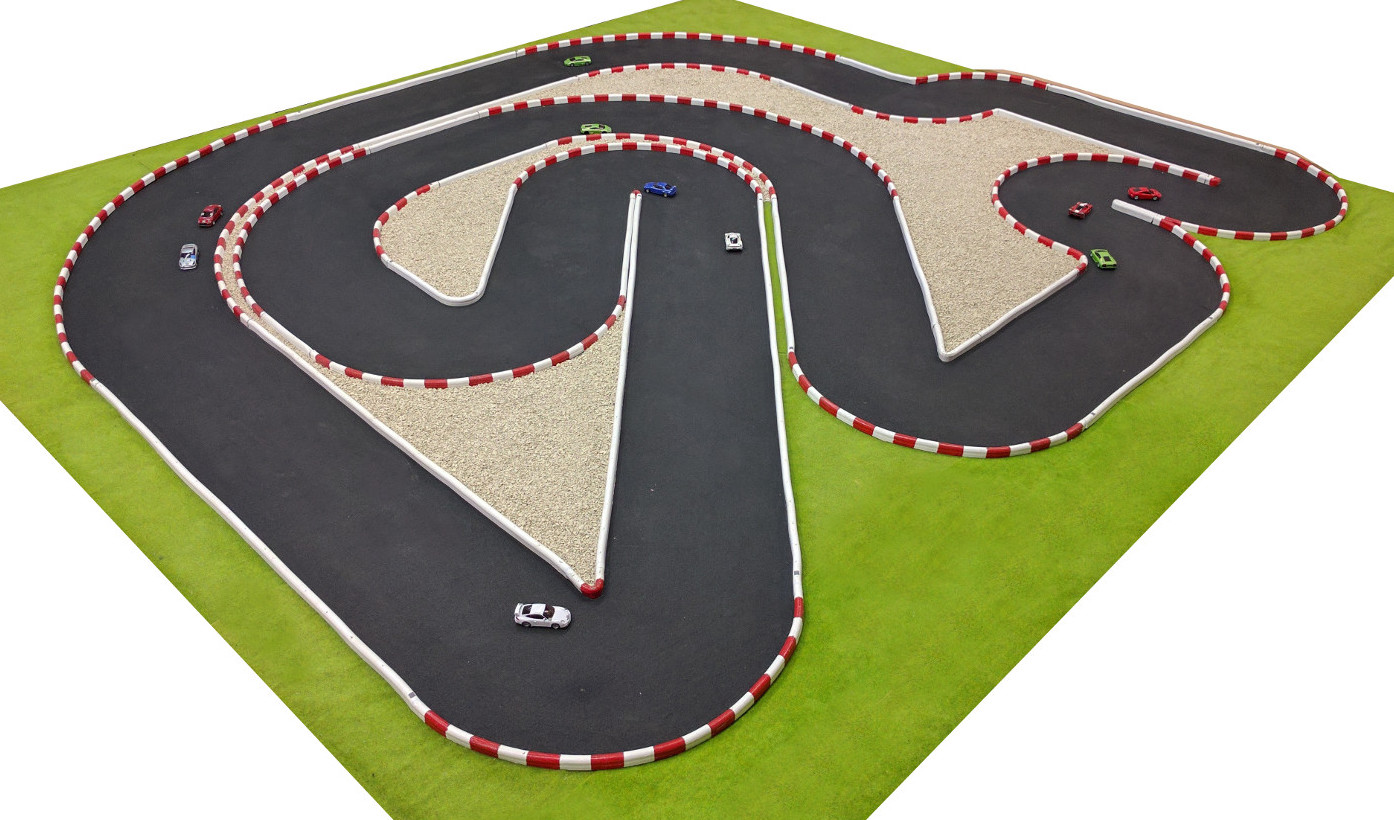}
  \caption{Picture of the track and Kyosho \emph{dnano} cars used in the experimental set up.}\label{fig:track}
\end{figure}

We consider a miniature race car driving along a track with known boundaries (Fig.\ \ref{fig:track}) and our autonomous racing controller is based on \cite{Liniger_2014}, where an optimization-based receding horizon control was proposed. The two level hierarchical structure of \cite{Liniger_2014} is designed such that, in the higher level, many possible finite horizon trajectories are generated based on the current state, track layout and a simplified car model. In the lower level, the trajectory with the largest progress is tracked using a MPC subject to the full car dynamics. In the following section we review the proposed controller and outline how viability theory may improve the implementation of the higher level. The material of this section is based on \cite{Liniger_2014} and \cite{Liniger_2015}.

\subsection{Vehicle model}
\begin{figure}[h]
\centering
\includegraphics[width = 0.65\textwidth]{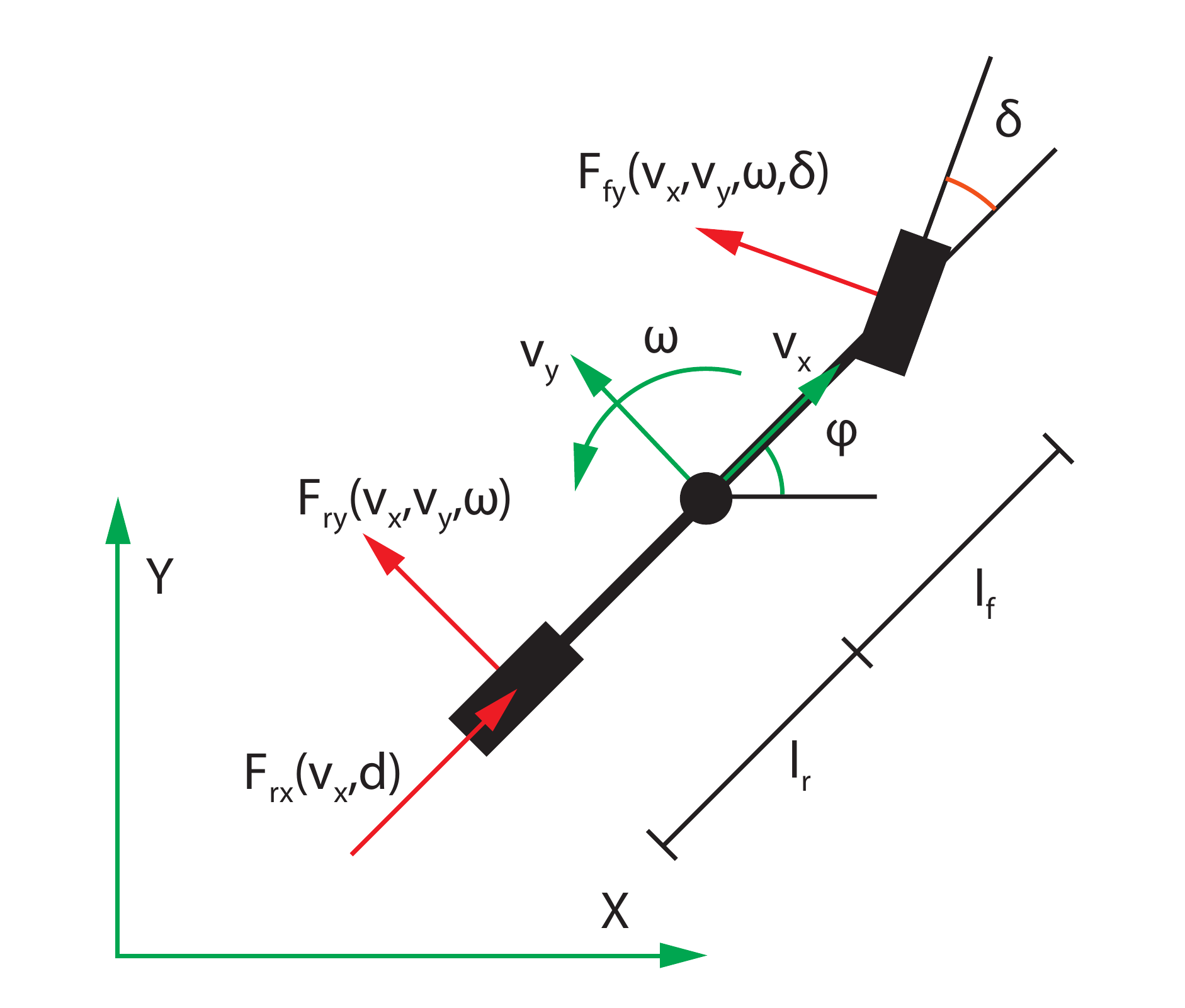}
  \caption{Schematic drawing of the car model}\label{fig:bicycle}
\end{figure}

The lower level control design is based on a nonlinear bicycle model (Fig. \ref{fig:bicycle}) using the Pacejka tire model \cite{MF}. This model captures important dynamics, such as the saturation of the nonlinear tire force, and the steering behavior at different forward velocities.

Following the literature, the equations of motion are derived around the center of gravity (CoG), and the states are the positions X and Y, and the orientation $\varphi$ relative to the inertial frame (Fig.\ \ref{fig:bicycle}). These three states characterize the kinematic part of the model. The second part of the model is derived in a body fixed frame centered at the CoG. The states are the longitudinal and lateral velocities $v_x$ and $v_y$ as well as the yaw rate $\omega$. The control inputs are the steering angle $\delta$ and the pulse width modulation duty cycle $d$ of the drive train motor. The complete equations of motion are
	\begin{align}
	\dot{X} &= v_x \cos(\varphi) - v_y \sin(\varphi)\,, \nonumber \\
	\dot{Y} &= v_x \sin(\varphi) + v_y \cos(\varphi)\,, \nonumber\\
	\dot{\varphi} &= \omega\,,\label{eq:odeModel}\\
	\dot{v}_x &= \frac{1}{m}\Big(F_{r,x}(v_x,d) - F_{f,y}(v_x,v_y,\omega,\delta) \sin{\delta} + m v_y \omega \Big)\,, \nonumber\\
	\dot{v}_y &= \frac{1}{m} \Big( F_{r,y}(v_x,v_y,\omega) + F_{f,y}(v_x,v_y,\omega,\delta) \cos{\delta} - m v_x \omega \Big) \,,\nonumber\\
	\dot{\omega} &= \frac{1}{I_z}\Big(F_{f,y}(v_x,v_y,\omega,\delta) l_f \cos{\delta} - F_{r,y}(v_x,v_y,\omega) l_r \Big)\,,\nonumber
	\end{align}
where $m$ is the mass of the vehicle, $l_r$ and $l_f$ are the distances from the CoG to the rear and the front wheel, respectively, and $I_z$ is the moment of inertia. $F_{r,x}(v_x,d)$ is the force produced by the drive train, $F_{r,y}(v_x,v_y,\omega)$ and $F_{f,y}(v_x,v_y,\omega,\delta)$ are the lateral forces at the rear and the front wheel, given by the Pacejka tire model \cite{MF}. For more details on the modeling as well as the exact tire model formulation we refer to \cite{Liniger_2014}.
\subsection{Path planning based on constant velocities} \label{subsec:pp}
As the model \eqref{eq:odeModel} is complex and has a fast changing dynamic, the higher level of the hierarchical controller, which we call the ``path planning controller", uses a simplified model to determine an optimal trajectory. Unlike most of the path planning algorithms in the literature, e.g. \cite{Frazzoli2005}, the goal of our path planning controller is not to reach a target while avoiding collision, but to find a finite horizon trajectory that stays within the track and which maximizes progress.

The bicycle model \eqref{eq:odeModel} is simplified by only considering trajectories that consist of segments of duration $T_{pp}$ with constant velocities. In vehicle dynamics such trajectories are known as \emph{steady state cornering}, whereas in aeronautics they are referred to as {\em trimmed flight}. The segments are then directly linked under the assumption that the new velocity can be reached immediately. To ensure that the low-level controller can track trajectories with velocity discontinuities, constraints are imposed on the segments that can be concatenated. This approach is motivated by the time scale separation present in the system where a system's velocity changes significantly faster than its position and orientation. A similar approach was adopted in \cite{Kitsios2005a,Tomlin98,Panagou2009} to reduce the number of states of the model. In our case, this approximation leads to a system with only four states, the three physical states $X$,$Y$ and $\varphi$ and a discrete state representing the constant velocity points, instead of the six states in equation \eqref{eq:odeModel}.

We calculate constant velocity points by fixing the steering angle and the longitudinal velocity, and determine the remaining velocities and inputs $(v_y, \omega, d)$ such that the accelerations of the model are zero. This corresponds to gridding the manifold describing the stationary velocities (Fig. \ref{fig:StatVeloMesh}) to generate a grid of $N_m$ constant velocity points, see Appendix \ref{app:N_m}.

\begin{figure}[h]
\centering
\includegraphics[width = 0.65\textwidth]{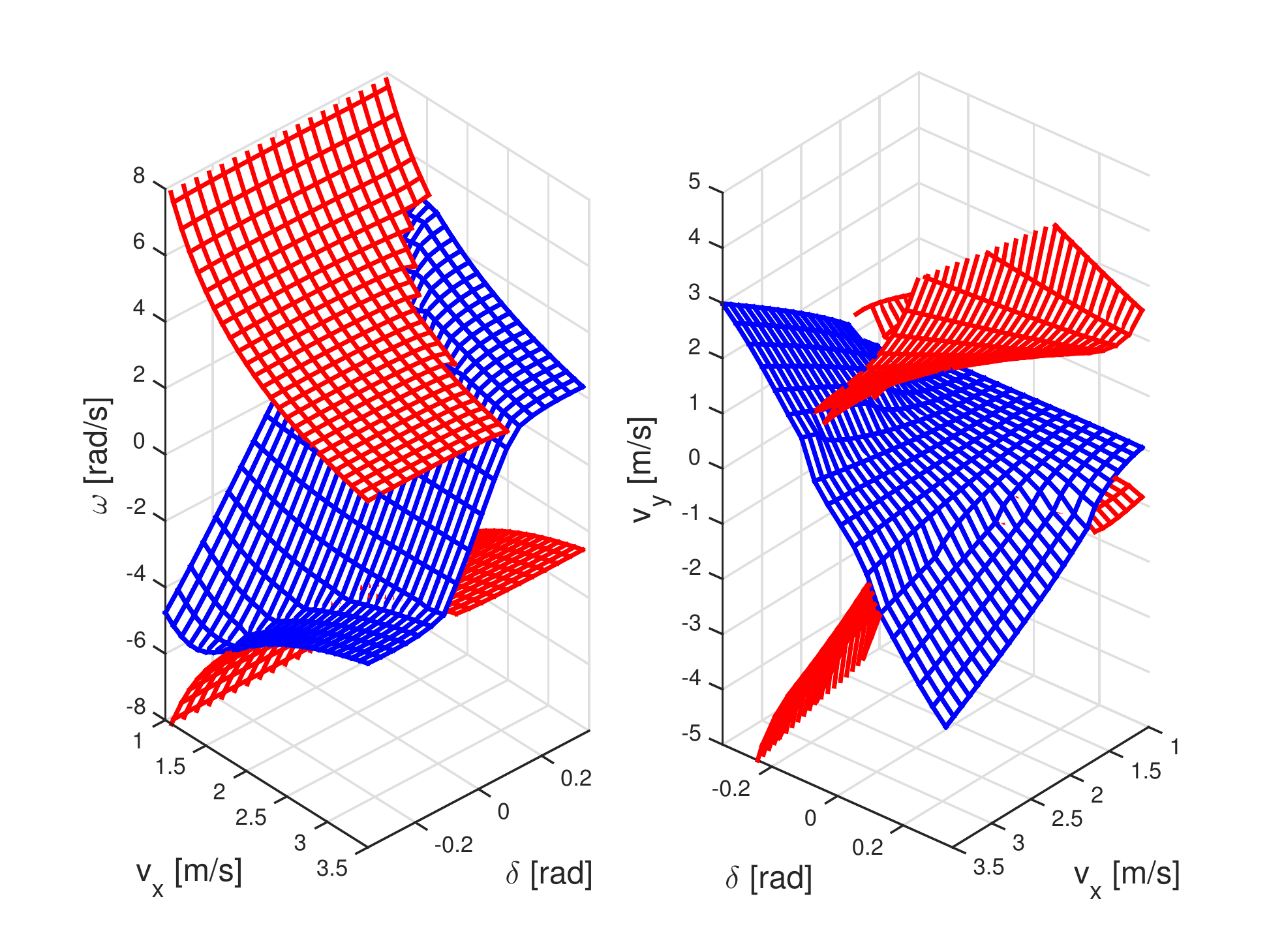}
  \caption{Stationary velocity manifold of the bicycle model rendered in red the parts corresponding to oversteering, in blue neutral and understeering, see \cite{Liniger_2014} for more details.}\label{fig:StatVeloMesh}
\end{figure}

To ensure that the velocities in concatenated segments are not unrealistic, we only allow for transitions that are achievable by the nonlinear dynamical model ($v_x$, $v_y$ and $\omega$) in a certain small time step $T_t \approx T_{pp}/2$. This problem can be posed as a feasibility problem where the goal is to find control inputs that allow the transition from one velocity to another within $T_t$ seconds. This allows one to find all admissible transitions between two stationary points. Note that the resulting concatenation constraints can be encoded through a finite state automaton and that together with the continuous dynamics of \eqref{eq:odeModel} this implies that the path planning problem can be cast as a hybrid control problem\cite{Liniger_2015}[Section 3].

\begin{figure}[h]
\centering
\includegraphics[trim = 5mm 10mm 5mm 25mm, clip, width = 0.65\textwidth]{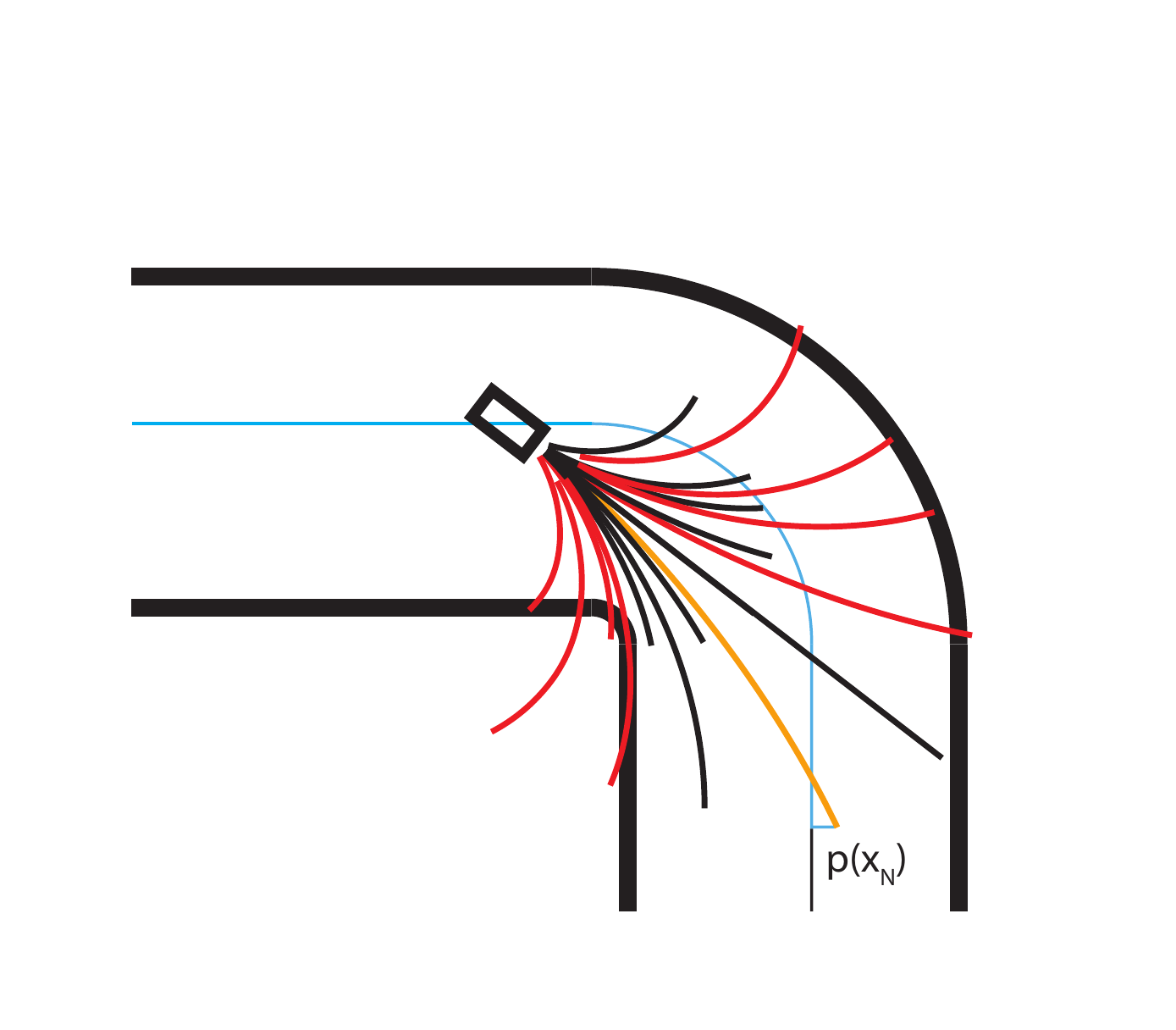}
  \caption{Schematic drawing of the path planning as presented in \cite{Liniger_2014}. The red trajectories get discarded as they leave the track; of the remaining trajectories (orange and black) the orange is the one leading to the greatest progress along the track indicated in blue.}\label{fig:PathPlanning}
\end{figure}

The path planner generates all trajectories that consist of $N_s$ constant velocity segments, each of duration $T_{pp}$, that satisfy the aforedescribed concatenation constraints. Once all the state trajectories have been computed the path planning algorithm discards those that violate the state constraints (i.e., leave the track). Among the remaining feasible trajectories, the path planner selects the trajectory with the greatest progress, calculated by projecting the end point of the trajectory on the center line of the track (Fig. \ref{fig:PathPlanning}). This is accomplished by first generating offline a piecewise affine approximation of the center line, comprising 488 pieces with end points distributed regularly on the center line. Online the projection can then be computed by finding the closest affine piece and taking an inner product.

A similar path planning concept based on constant velocity segments (trims) was derived in \cite{Frazzoli2005}, and applied to autonomous driving in \cite{Gray2012}. Our method is different since we do not plan a path from a starting point to a target set.

\subsection{Reference tracking MPC}
The selected trajectory from the higher level path planner is passed over to a lower level MPC that tracks the given trajectory based on the full model \eqref{eq:odeModel}. To render the MPC problem tractable, first the nonlinear model \eqref{eq:odeModel} is linearized around the reference state and input trajectory from the path planner and then discretized. Second, the track constraints are approximated by constraining each $X-Y$ state along the horizon to lie between two half-spaces approximating the track. These reformulations, together with a quadratic state reference tracking cost and a quadratic input rate cost, allow to formulate the problem as a convex quadratic program. The resulting optimization problem can be solved within milliseconds using tailored MPC solvers such as \cite{Domahidi2012}. Since the focus of the paper is on the higher level path planner, we refer the interested reader to \cite{Liniger_2014}, where the MPC problem is discussed in more detail.

\subsection{Limitations and motivations of present work} \label{subsec:limitations}
While the hierarchical controller outlined above has enabled us to perform real-time racing of the RC cars, it faces two major drawbacks that limits its performance. The first drawback is the exponential growth of possible trajectories as a function of the number of segments, $N_s$. The second drawback is that feasibility does not imply recursive feasibility. Indeed it has been observed in \cite{Liniger_2015}, that the maximal progress trajectory is often not recursive feasible, when only feasibility is enforced.

To tackle these drawbacks we calculate the viability kernel of the path planner. By restricting the path planner to trajectories that keep the system within the viability kernel (instead of just the track), recursive feasibility of the path planning step is guaranteed under the assumption of no model mismatch. Moreover, the use of the viability kernel has two computational advantages that can be exploited to dramatically reduce the computation time. First, it allows one to consider fewer trajectories (only those that remain in the viability kernel instead of all trajectories that remain in the track) and second, it suffices to verify that at any switching point stays within the viability kernel instead of checking that the whole trajectory stays within the track. Note that using the viability kernel does not fundamentally tackle the exponential growth of the path planning computation in the horizon length. Empirically, however, this additional pruning of candidate trajectories does lead to a significant reduction in the online computation times. Preliminary results reported in \cite{Liniger_2015} suggest that this reduction is by a factor of ten, making it possible to extend the paths considered by more segments, leading to a better closed-loop performance. In Section \ref{sec:sim} we show that a more efficient implementation allows a further reduction in computation time and allow for even more segments.

\section{Viability kernel and discriminating kernel} \label{sec:Viab}
In this section we briefly discuss discrete-time viability theory which will later on be used to construct recursive feasible trajectories. Discrete-time viability theory addresses the question for which initial conditions does there exists a solution to a difference inclusion, which stays within a constraint set forever \cite{Aubin2009}. Consider a controlled discrete-time system $x_{k+1} = f(x_k,u_k)$, where $x \in \mathbb{R}^n$ is the state, $u \in U \subset \mathbb{R}^m$ is the control input and $f:\mathbb{R}^n\times U \rightarrow \mathbb{R}^n$ a continuous function. This system can be formulated as a difference inclusion,
\begin{align}
x_{k+1} \in F(x_k), \qquad \text{with } F(x) =  \{f(x,u)| u \in U\}\,. \label{eq:diffInclusion}
\end{align}
We next briefly summarize the standard viability kernel algorithm of \cite{saintPierre94} for the above class of systems.
\subsection{Viability kernel algorithm}
Given a constraint set $K\subset \mathbb{R}^n$, solutions to the difference inclusion \eqref{eq:diffInclusion} which stay in $K$ forever are known as \emph{viable solutions}.
\begin{definition}[\hspace*{-1.7mm}\cite{saintPierre94}]
A set $D \subset \mathbb{R}^n$ is a \textbf{discrete viability domain} of $F$ if $F(x) \cap D \neq \emptyset$ for all $x \in D$. The \textbf{discrete viability kernel} of a set $K \subset \mathbb{R}^n$ under $F$, denoted by $\Viab_F(K)$, is the largest closed discrete viability domain contained in $K$.
\end{definition}
Under mild assumptions on $F$ (discussed below in Proposition \ref{th:disViabAlge}) standard viability theory arguments ensure the existence of the discrete viability kernel and establish the existence of viable solutions for all the states contained in it.

Conceptually speaking, the viability kernel can be calculated through the so called viability kernel algorithm:
\begin{align}
&K^0 = K\,, \nonumber\\
&K^{n+1} = \{x \in K^n | F(x) \cap K^n \neq \emptyset\}\,. \label{eq:viabAlgo}
\end{align}
\begin{proposition}[\hspace*{-1.2mm}\cite{saintPierre94}] \label{th:disViabAlge}
Let $F$ be an upper-semicontinuous set-valued map with closed values and let $K$ be a closed subset of $\textnormal{Dom}(F)$. Then, $\bigcap_{n=0}^{\infty} K^n = \Viab_F(K)$.
\end{proposition}
The viability kernel algorithm requires one to perform operations with arbitrary sets and as such is not implementable. In practice, viability kernels are approximated by discretizing/gridding the state space. Consider a family of countable subsets $X_h$ of $\mathbb{R}^n$, parameterized by $h \in \mathbb{R}$, such that
\begin{align*}
\forall x \in \mathbb{R}^n , \quad \exists x_h \in X_h \quad \text{s.t.} \; \| x-x_h\|_{\infty} \leq \alpha(h)\,,
\end{align*}
for some continuous function $\alpha: \mathbb{R}_+ \rightarrow \mathbb{R}_+$ with $\lim_{h \rightarrow 0} \alpha(h) = 0$. Given a grid point $x_h \in X_h$, we will call the set of points $\{x \in \mathbb{R}^n \; | \; \|x-x_h\|_\infty \leq \alpha(h)\}$ the \emph{cell} of $x_h$. Notice that the cells of different grid points may overlap. We note that, unlike \cite{saintPierre94}, we resort here to the infinity norm, as this will facilitate the presentation of subsequent results. 

To ensure that the discretized set-valued map $F(x_h) \cap X_h$ is non-empty it is necessary to enlarge the set-valued map before discretization, a process that is known as ``expansion" \cite{saintPierre94}. The expanded set-valued map is defined as $F^r(x) = F(x) + r B_{\infty}$, where $B_{\infty}$ is a closed unit ball in the infinity norm centered at the origin and $r$ is the radius of the ball. It can be shown that, for $r \geq \alpha(h)$,  $F^r(x) \cap X_h$ is non empty for all $x$. Throughout we denote $F_h^r(x_h) = F^r(x_h) \cap X_h$ and assume for simplicity that $r=\alpha(h)$. When restricted to $X_h$ the set-valued map $F_h^r$ defines the following discrete time system over the countable set by
\begin{align} 
x_{h,k+1} \in F_h^r(x_{h,k})\,. \label{eq:finitDynSys}
\end{align}
If the set $K$ is compact, then the set $X_h$ can be assumed to be finite and the viability kernel algorithm \eqref{eq:viabAlgo} applied to \eqref{eq:finitDynSys} converges after a finite number of iterations, since $K_h = K \cap X_h$ is finite. Moreover, under appropriate technical assumptions \cite{saintPierre94}, the resulting set inner-approximates the viability kernel, i.e.,
\begin{align}
\Viab_{F_h^{r}}(K_h) \subset (\Viab_{F^r}(K) \cap X_h) \,. \label{eq:finiteGuarantee}
\end{align}
Note that the guaranteed inner approximation is only with respect to the viability kernel of the extended dynamics $F^r$. The extension limits the granularity/resolution of features that the kernel can represent due to extension, but is necessary in the finite case. Finally, the finite viability kernel converges to the discrete viability kernel as the space discretization $h$  goes to zero, \cite{saintPierre94}.
\subsection{Discriminating kernel algorithm}
The discriminating kernel algorithm of\cite{Cardaliaguet99} extends viability computations to systems where the state evolution $x_{k+1}=g(x_k,u_k,v_k)$ is additionally affected by a disturbance $v_k \in V \subset \mathbb{R}^d$, where $g:\mathbb{R}^n \times U \times V \rightarrow \mathbb{R}^n$ is assumed to be a continuous function. The discriminating kernel algorithm returns all states (``victory domain") where there exists a control input, which is able to prevent the disturbance from driving the system to the open set $\mathbb{R}^n\setminus K$. Notice that, in the discriminating kernel the control is able to access the disturbance's current action, which is not the case if the related leadership kernel is used, for more details we refer to \cite{Cardaliaguet1996}.

The difference equation can again be written as a difference inclusion 
\begin{align*}
x_{k+1} \in G(x,v) \quad \text{with } G(x,v) = \{g(x,u,v) | u \in U\}\,.  
\end{align*}
\begin{definition}[\hspace*{-1.mm}\cite{Cardaliaguet99}]
 A set $Q \subset \mathbb{R}^n$ is a \textbf{discrete discriminating domain} of $G$, if for all $x \in Q$, we have that $G(x,v) \cap Q \neq \emptyset$ for all $v \in V$. The \textbf{discrete discriminating kernel} of a set $K \subset \mathbb{R}^n$ under $G$, denoted by $\Disc_G(K)$, is the largest closed discrete discriminating domain contained in $K$.
\end{definition}
One can show that the discriminating kernel  $\Disc_G(K)$ exists if $K$ is a closed subset of $\mathbb{R}^n$ and $G:\mathbb{R}^n\times V \rightsquigarrow \mathbb{R}^n$ is an upper-semicontinuous set-valued map with compact values \cite{Cardaliaguet99}.
The discrete discriminating kernel can be calculated using an algorithm similar to the viability kernel algorithm, by considering a sequence of nested closed sets
\begin{align}
&K^0 = K\,, \nonumber\\
&K^{n+1} = \{x \in K^n | \forall v\in V, \; G(x,v) \cap K^n \neq \emptyset  \}\,.  \label{eq:discAlgo}
\end{align}
To numerically implement the discriminating kernel algorithm, the state space $\mathbb{R}^n$, the disturbance space $V$ and the difference inclusion $G(x,v)$ have to be discretized \cite{Cardaliaguet99}. Whenever $K$ and $V$ are compact the algorithm \eqref{eq:discAlgo} terminates after a finite number of iterations. Similar to the viability kernel it can be shown that the finite approximation of the discriminating kernel converges to the discrete discriminating kernel as the space discretization $h$ goes to zero \cite{Cardaliaguet99}.

\section{Robustifying the viability kernel algorithm against space discretization errors} \label{sec:spaceDis}
\subsection{Errors due to the space discretization} 
We start by noting that the inner approximation property of the viability kernel algorithm \eqref{eq:finiteGuarantee} only holds for viable grid points and not for states in their cells. In practice it is desirable that the viability property of a grid point is passed on to its cell. In the following section we present a rigorous approach to ensure this property.

\begin{assumption} \label{as:Lipschitz}
The function $f(x,u)$ is continuous in $u$ and globally Lipschitz in $x$ with Lipschitz constant $L$, that is there exists a positive number $L$, such that for all $u \in U$,
\begin{align*}
\forall x,y \in \mathbb{R}^n : \| f(y,u) - f(x,u)\|_{\infty} \leq L\| y - x\|_{\infty}\,.\\[-0.3cm] \nonumber
\end{align*}
\end{assumption}
Consider $x_h \in X_h$, and a point $x \in x_h + rB_\infty$ in the cell of $x_h$, then under Assumption \ref{as:Lipschitz}, it holds that for all $u \in U$ 
\begin{align}
f(x,u) \in f(x_h,u)  + L r B_{\infty}\,.\label{eq:LipschitzBox3}
\end{align}
Therefore, to robustify the viability kernel algorithm against the space discretization, we can model the discretization error as an additive disturbance on our nominal system, i.e.,
\begin{align}
x_{k+1} = f(x_k,u_k)  +  v_k, \qquad \mbox{with } v_k\in L r B_{\infty}\,. \label{eq:addUncertainty}
\end{align}
\subsection{Inner approximation of viability kernel} \label{sec:4}
To calculate an inner approximation of the viability kernel that guarantees viability of the entire cell of a viable point, we propose to use the discriminating kernel algorithm \eqref{eq:discAlgo}. 

\begin{proposition} \label{prop:PropertiesAlgo}
Under Assumption \ref{as:Lipschitz}, the discriminating kernel $\Disc_G(K)$ of the difference inclusion,
\begin{align} \label{eq:dynSysOfInterest}
x_{k+1} \in G(x_k,v_k) =  F(x_k) + v_k \quad v_k\in L r B_{\infty}\,,
\end{align}
enjoys the following properties:
\begin{itemize}
\item[1)] $\lim_{r \rightarrow 0} \Disc_{G}(K) = \Viab_F(K)$.
\item[2)] $\Disc_G(K)$ is a viability domain of $F$.
\item[3)] For all $x \in \Disc_G(K)$ and for all $\hat{x} \in x +r B_\infty$, there exists a $u \in U$ such that $f(\hat{x},u) \in \Disc_G(K)$.
\end{itemize}
\end{proposition}
\begin{IEEEproof} Under Assumption \ref{as:Lipschitz} $F(x)$ is upper-semicontinuous, hence $G(x)$ is upper-semicontinuous, and the conditions to apply the viability and discriminating kernel algorithms are satisfied.  \newline
1) If $r \rightarrow 0$ the disturbance set vanishes, thus $\bigcap_{r>0} L r B_{\infty} = \{ 0 \}\ $
and using \emph{Remark 4.1} of \cite{Cardaliaguet99}, the discriminating kernel is identical to the viability kernel. \newline
2) By definition the discriminating kernel is a discriminating domain.
Thus for all $x \in \Disc_G(K)$ we have,
\begin{align*}
(F(x) + v) \cap \Disc_G(K) \neq \emptyset \quad \forall v \in V =L r B_{\infty}\,,
\end{align*}
Since $v=0 \in LrB_\infty$ is an allowed disturbance, the discriminating kernel is also a viability domain, as
\begin{align*}
\forall x \in \Disc_G(K), \quad F(x) \cap \Disc_G(K) \neq \emptyset\,.
\end{align*}
3) First, by \eqref{eq:LipschitzBox3} we know that for all $\hat{x} \in x +r B_\infty$
\begin{align*}
f(\hat{x},u) \in f(x,u) + LrB_\infty\,,
\end{align*}
and therefore for a given $u\in U$, $f(\hat{x},u) = f(x,u) + v$, for some $ v \in LrB_\infty$.
Second, from \cite{Cardaliaguet1994}[Equation (4)] we know that the discriminating kernel $\Disc_G(K)$ is the largest closest subset of $K$ such that for all $x \in \Disc_G(K)$ and for all $v \in V$, there exists a $u \in U$ such that $g(x,u,v) \in \Disc_G(K)$.
Thus, together we have that for every $\hat{x} \in x +r B_\infty$ there exists a $u\in U$, such that $f(\hat{x},u) \in \Disc_G(K)$.
\end{IEEEproof}

Note that if the leadership kernel is used instead of the discriminating kernel, a similar property can be shown.

As discussed in the previous Section \ref{sec:Viab}, it is necessary to discretize space to implement the discriminating kernel algorithm. To state similar properties as in Proposition \ref{prop:PropertiesAlgo} if the space is discretized we need an inner approximation of the discriminating kernel $\Disc_{G^r_h}(K_h) \subset \Disc_{G^r}(K)\cap X_h$. This is generally not given, therefore we derive a modified discriminating kernel algorithm, which guarantees an inner approximation for the given system \eqref{eq:dynSysOfInterest}. The modified discriminating kernel algorithm is derived in Appendix \ref{app:InnerApprox} and stated in Algorithm \ref{algo:innerApproxDisc}. The algorithm, however needs two assumptions; $(i)$ we only consider square grids and $(ii)$ the set $U$ needs to be finite. The second assumption is necessary to implement the algorithm and can be achieved by introducing a finite discretization of $U$, which we henceforth denote by $U_h \subset U$. Note that for our system square grids are an obvious simple choice and the path planning model has a finite input space by construction.
\begin{proposition} \label{prop:controlInnerApprox}
Consider the finite dynamical system corresponding to the extended and discretized system of \eqref{eq:dynSysOfInterest},
\begin{align*}
x_{h,k+1} \in G^r_h(x_{h,k}) =  (F(x_{h,k}) + v_h + r B_\infty)\cap X_h\,, 
\end{align*}
where $v_h \in V_h$ is the discretization of $L r B_\infty$. If $\Disc_{G^r_h}(K_h)$ is computed with Algorithm \ref{algo:innerApproxDisc}, then the following properties hold:
\begin{itemize}
\item[1)] $\Disc_{G^r_h}(K_h)$ is a viability domain of $F^r_h$.
\item[2)] For all $x_h \in \Disc_{{G}^r_h}(K_h)$ and for all $\hat{x} \in x_h + rB_\infty$, there exists $u_h \in U_h$ such that $(f(\hat{x},u_h) + rB_\infty)\cap X_h \in \Disc_{{G}^r_h}(K_h)$.
\end{itemize}
\end{proposition}
\begin{IEEEproof} 
See Appendix \ref{app:InnerApprox}.
\end{IEEEproof}

\subsection{Reconstructing viable controls} \label{sec:6}
The use of the discriminating kernel in the path planning step of the hierarchical racing controller is not directly possible as the action depends on both the state and the disturbance. Thus, the reconstruction of a viable set-valued feedback policy as proposed in \cite{Liniger_2015} is not possible. However, as the disturbance is related to the position of the current state within the grid cell the disturbance is indirectly known. Furthermore, we know that if the state is within a grid cell of the discriminating kernel, there exists a control which keeps the system within a grid cell of the discriminating kernel, see Proposition \ref{prop:controlInnerApprox}. Thus, we can formulate a predictive controller which enforces this property. More precisely, we can consider the following optimization problem which constrains the state to stay within the discriminating kernel. Feasibility of the problem below is ensured by virtue of Proposition \ref{prop:controlInnerApprox}; the objective function $J(x,u)$ allows us to optimize some cost function. In our autonomous racing application $J(\cdot,\cdot)$ is chosen such that the car's progress, measured as described in Section \ref{sec:HRHC}, is maximized, and the system dynamics $x_{k+1} = f(x_k,u_{h,k})$ is the path planning model described in the next section,
\begin{align} 
&\min_{{\mathbf{u}_h,\mathbf{x}}} \quad \sum_{k=0}^{N_S} J(x_k,u_{h,k})\,,\nonumber \\ \label{eq:PredController}
&\; \text{ s.t.} \; \quad x_0 = x\,,\\ 
& \qquad \quad x_{k+1} = f(x_k,u_{h,k})\,, \quad u_{h,k} \in U_h\,, \nonumber \\ 
& \qquad \quad (f(x_k,u_{h,k}) + rB_\infty)\cap X_h \subset \Disc_{G_h^{r}}(K_h)\,. \nonumber
\end{align}
Where, $\mathbf{u}_{h} = [u_{h,0},...,u_{h,N_S-1}]$ is the control sequence, $\mathbf{x} = [x_{0},...,x_{N_S}]$ the corresponding state sequence, $x$ the measured state and $u_{h,k} \in U_h$ the input constraints. We point out that the problem is solved by enumeration, which limits the horizon to few steps. To reduce the computation time related to generate all feasible input and state sequences, it is possible to replace $u_{h,k} \in U_h$ with $u_{h,k} \in U_D(x_k +rB_\infty \cap X_h)$, where, 
\begin{align}
U_{\text{D}}(x_h)  = \begin{cases}
\hspace*{-1mm}\begin{Bmatrix}
u_h \in U_h\, |\exists v_h \in V_h \\
f(x_h,u_h) +v_h+r B_\infty\\ 
 \cap  \Disc_{G_h^{r}}(K_h) \neq \emptyset
 \end{Bmatrix}
 &\begin{array}{l} \hspace*{-4mm}  \text{if } x_h \in \\ \hspace*{-4mm}  \Disc_{G_h^{r}}(K_h) \end{array} \\ \\
\quad \emptyset & \hspace*{-2mm}\text{otherwise}
\end{cases} \hspace*{-2mm}. \label{eq:safeInput}
\end{align}
If instead the viability kernel is used, the last constraint of \eqref{eq:PredController} is replaced with $(f(x_k,u_{h,k}) + rB_\infty)\cap X_h \subset \Viab_{F_h^{r}}(K_h)$. And to reduce the computation time $u_{h,k} \in U_h$ can be replaced with $u_{h,k} \in U_V(x_k+rB_\infty \cap X_h)$, where the set-valued feedback law $u_{k,h} \in U_{\text{V}}(x_h)$ can be computed similar to $U_{\text{D}}$ in \eqref{eq:safeInput}. However, there exists no guarantee that the control will keep the system within the viability kernel if the state does not always coincide with a grid point.

In the following two sections, we discuss $x_{k+1} = f(x_k,u_k)$ as well as $\Disc_G(K)$ and $\Viab_F(K)$, which are needed to formulate the high level path planner in terms of \eqref{eq:PredController}.

\section{Path planning model} \label{sec:Hybrid}
As discussed in Section \ref{sec:HRHC}, the path planning problem can be formalized in terms of hybrid systems, where the  discrete mode $q$ of the hybrid system represents the current stationary velocity, while the continuous evolution is given by the kinematic model with the corresponding stationary velocity. A clock variable is added which allows a jump to a different discrete mode every $T_{pp}$ seconds. Thus the continuous evolution is given by,
\begin{subequations}\label{eq:contEvolution}
\begin{align} 
\dot{X} &= v_x(q) \cos(\varphi) - v_y(q) \sin(\varphi)\,, \label{eq:contEvolution1} \\
\dot{Y} &= v_x(q) \sin(\varphi) + v_y(q) \cos(\varphi) \,,\label{eq:contEvolution2} \\
\dot{\varphi} &= \omega(q)\,,\label{eq:contEvolution3}\\
\dot{T} & = 1\,,
\end{align}
\end{subequations}
where $v_x(q), v_y(q), \omega(q)$ are the stationary velocities at the discrete mode $q$. The discrete transition takes place every $T_{pp}$ seconds, and a discrete control input respecting the concatenation constraints encoded by the finite automaton determines the discrete mode after the transition.
As shown in \cite{Liniger_2015}, the model can be formally cast in the framework of hybrid automata \cite{Lygeros2003}.

Since the system is hybrid one would, strictly speaking, have to employ hybrid viability algorithms \cite{Aubin2002,Margellos2013}. This would, however, require gridding the four dimensional continuous space for each of the $N_m$ discrete modes, a task that is computationally very demanding. As the discrete transitions occur at fixed time intervals, the system can be interpreted as a sampled data system, with constant inputs over the time interval. Therefore, the system can be approximated by a  discrete time system with a sampling time of $T_{pp}$. The discrete state $q$ of the hybrid system is considered by embedding it in the real numbers, or in other words ``pretending" that $q \in \mathbb{R}$; hence the discrete dynamics can be encoded as a transition relation that depends on the current mode. This renders the time state redundant and reduces the state space dimension by one. Thus, it is only required to grid a 3 dimensional space, in addition to the coarse grid corresponding to the discrete state.

We denote the state space of the difference equation at time step $k$ by $x_k = (X_k,Y_k,\varphi_k,q_k)$, and the control by $u_k$, which determines the next mode $q_{k+1} = u_k$. As the control input depends on the current mode $q_k$, we abstractly write $u_k \in U(x_k)$, where $U(x_k)$ represents all allowed transitions in the automaton encoding the concatenation constraints, or in other words all allowed next modes $q_{k+1}$, as detailed in \cite{Liniger_2015}. Let  $X(x_k,u_k,t),\; Y(x_k,u_k,t),\; \varphi(x_k,u_k,t)$ denote the solution of \eqref{eq:contEvolution1}-\eqref{eq:contEvolution3} at time $t$, with $q = u_k$ while starting at $x_k = (X_k,Y_k,\varphi_k,q_k)$. Note that, since the velocities are constant the solution can be explicitly computed as a function of time. Thus, we can define the path planner as the following discrete-time model,
\begin{align} \label{eq:PPmodel}
x_{k+1} = f(x_k,u_k) &= \begin{bmatrix}
X(x_k,u_k,T_{pp})\\
Y(x_k,u_k,T_{pp} ) \\
\varphi(x_k,u_k,T_{pp})\\
u_k
\end{bmatrix}\,,
\end{align}
where $T_{pp}$ is the path planning discretization time. The set-valued map needed for the viability kernel algorithm is given by $F(x) = \{f(x,u)|u \in U(x)\}$, and the output is the union of a finite number of points in the state space. 

For the viability and discriminating kernel algorithm it is required that $F(x)$ is an upper-semicontinuous set-valued map. In the following, we show that by appropriately embedding the discrete state in the set of real numbers, upper-semicontinuity of the set-valued map of \eqref{eq:PPmodel} can be ensured despite the hybrid structure of the model. To see this, recall that a set-valued map $F:\mathbb{R}^n \rightsquigarrow \mathbb{R}^n$ is called upper-semicontinuous if and only if for all $x\in \mathbb{R}^n$ and for all $\epsilon > 0$ there exists an $\eta >0$ such that for all $x' \in x + \eta B$ it holds that $F(x') \subset F(x) + \epsilon B$ \cite{Aubin2009set}. By embedding the discrete state $q$ in the real numbers we have $x = (X,Y,\varphi,q) \in \mathbb{R}^4$, where the first three states $(X,Y,\varphi)$ are continuous by virtue of the original model. Collecting them into the continuous state $x_c = (X,Y,\varphi) \in \mathbb{R}^3$ and defining the integer lattice $\mathbb{M} = \{1,...,N_m\}$, where $N_m$ is the number of modes, we can define the following upper-semicontinuous set-valued map,
\begin{align*}
{F}(x_c,q) = \begin{cases}
\quad\tilde F(x_c,m), & \text{if } q \in (m-\frac{1}{2},m+\frac{1}{2}) \\[+0.3cm]
\begin{Bmatrix}\tilde F(x_c,m),\\ \tilde F(x_c,m+1)\end{Bmatrix},& \text{if }  q = m+\frac{1}{2} \\[+0.3cm]
\end{cases} \,,
\end{align*}
where $\tilde F(x_c,m) = \{f([x_c,m],u)| u \in U(m)\}$ is the set-valued map defined at the lattice point $m$. Notice that $q \in (1-1/2,N_m+1/2)$ is w.l.o.g. since the automaton maps the interval to itself, forming an invariant set in $\mathbb{R}$. Finally, upper-semicontinuity is ensured because the dynamics of $x_c$ is continuous and, using the above proposed formulation there, are no discontinuities in a set-valued sense.

To reduce the number of grid points in a practical implementation of the algorithm it is beneficial to limit the angle between $0$ and $2\pi$. Instead of using $\mathrm{mod}(\varphi,2\pi)$, which would lead to a discontinuous set-valued map, we propose the set-valued $\mathrm{mod}$ function
\begin{align*}
\mathrm{Mod}(\varphi,2\pi) = \begin{cases}
\mathrm{mod}(\varphi,2\pi) & \text{if } \varphi \neq k 2\pi \\
[0,2\pi] & \text{otherwise} \\
\end{cases} \,,
\end{align*}
which is upper-semicontinuous.

Finally, with the model \eqref{eq:PPmodel} it is still possible that the trajectory leaves the constraint set $K$ and re-enters during the time interval $T_{pp}$. This is an issue more generally for sampled data systems and was tackled in \cite{Mitchell2015} by a sampled data system viability/discriminating kernel algorithm that discards control inputs and states for which the continuous evolution of the data sampled system leaves the constraint set $K$. 

In the following section we will use the described model, including the modifications and the modified sampled data system viability/discriminating kernel algorithm to compute the kernel with respect to the track constraints.

\section{Viability kernel for the track} \label{sec:viabKernelTrack}
\subsection{Track constraints}
The goal is to find a viable solution of the path planning model within the track (Fig. \ref{fig:track}). To that end, we can define the constraint set $K$ as,
\begin{align*}
K := \left\{ \begin{array}{l}
(X,Y) \in \mathcal{X}_{\text{Track}}\,,\\
\varphi \in [0, 2\pi]\,,\\
q \in \{1,...,N_m\}\,.
 \end{array} \right.
\end{align*}
In other words, $X,Y$ are constrained within the track, while $\varphi$ and $q$ are unconstrained. The state space is uniformly gridded such that the distance in the respective unit (meter or radian) between two grid points is identical. This leads to the smallest $r$ if the states are not normalized. Further, as $q$ is already finite no gridding is necessary.

The Lipschitz constant of the path planning model \eqref{eq:PPmodel} can be calculated analytically by differentiating the explicit formula for $f(x_k,u_k)$. By using different Lipschitz constants for each mode it is possible to reduce conservatism of the proposed discriminating kernel approximation.

\subsection{Viability vs. discriminating kernel}
\begin{figure}[h]
\centering
\includegraphics[trim = 0mm 0mm 0mm 40mm, clip, width = 0.65\textwidth]{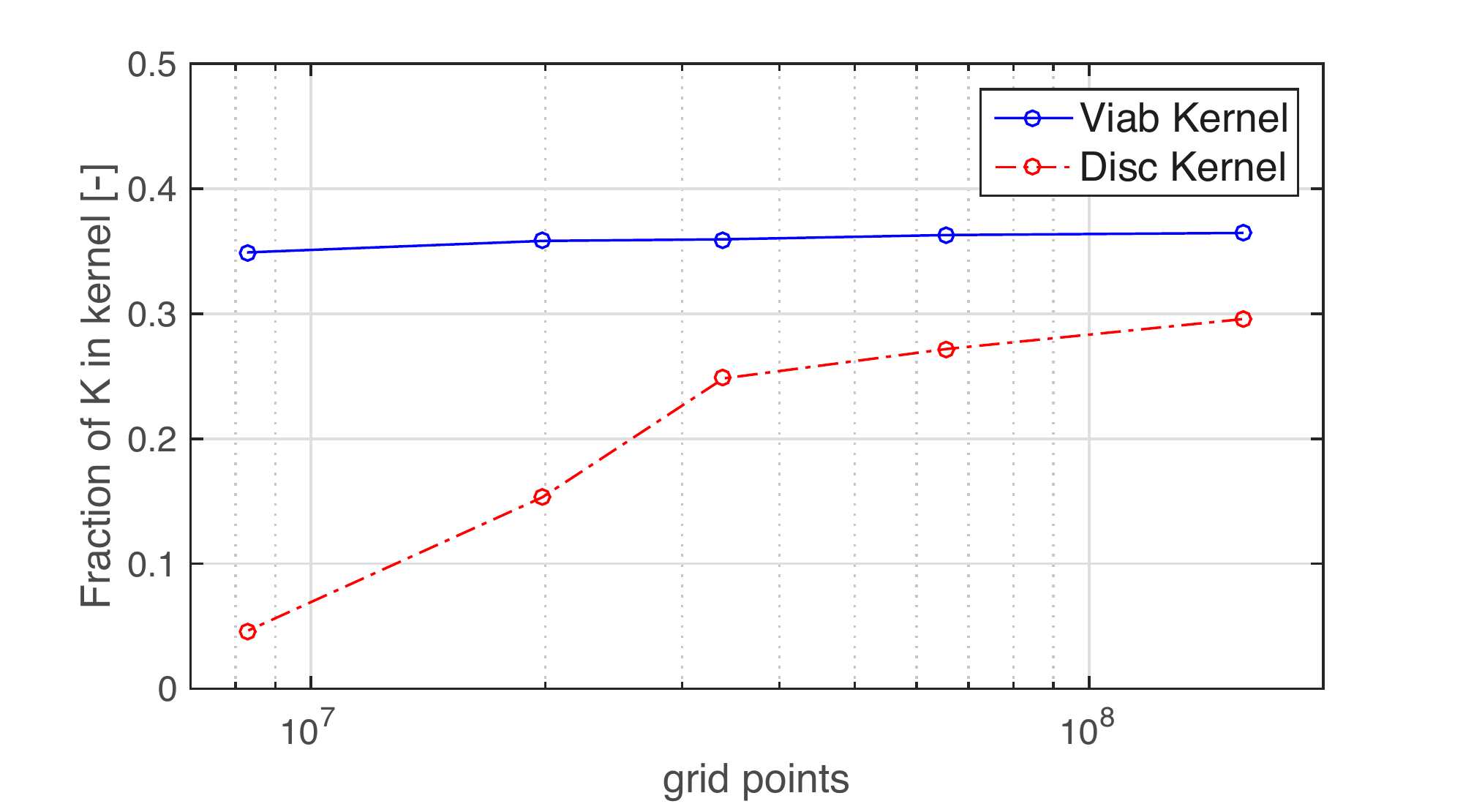}
  \caption{Fraction of grid points of $K$ which fall into the viability kernel and the discriminating kernel, as a function of the total number of grid points in $K$.}\label{fig:ViabVsDiscVol}
\end{figure}

Proposition \ref{prop:PropertiesAlgo} statement 1) in Section \ref{sec:spaceDis} suggests that the proposed discriminating kernel approximation converges to the viability kernel as the grid spacing $r$ goes to zero. To verify this, we compare the fraction of grid points in $K$ which fall into the respective kernels. Thus, if all grid points in $K$ are viable, then the fraction would be one. 
Fig. \ref{fig:ViabVsDiscVol} shows the fraction for the viability kernel and the discriminating kernel approximation for different numbers of grid points. It is interesting to see that the proportion of the viability kernel stays approximately at 0.35, whereas the discriminating kernel approaches the viability kernel as the grid spacing gets smaller. We can make two observations from Fig. \ref{fig:ViabVsDiscVol}, first, as Proposition \ref{prop:PropertiesAlgo} suggests the two approximations of the viability kernel get closer the smaller the grid spacing is. Second, while the viability kernel only changes slightly as a function of the grid spacing, the proposed discriminating kernel approximation requires a relative fine grid to achieve an approximation that is similar to the one of the viability kernel algorithm. Intuitively speaking, this is due to the fact that a coarse grid introduces a larger uncertainty than a fine grid, see also \eqref{eq:addUncertainty}.

\begin{figure}[h]
\centering
\includegraphics[trim = 5mm 20mm 5mm 25mm, clip, width = 0.65\textwidth]{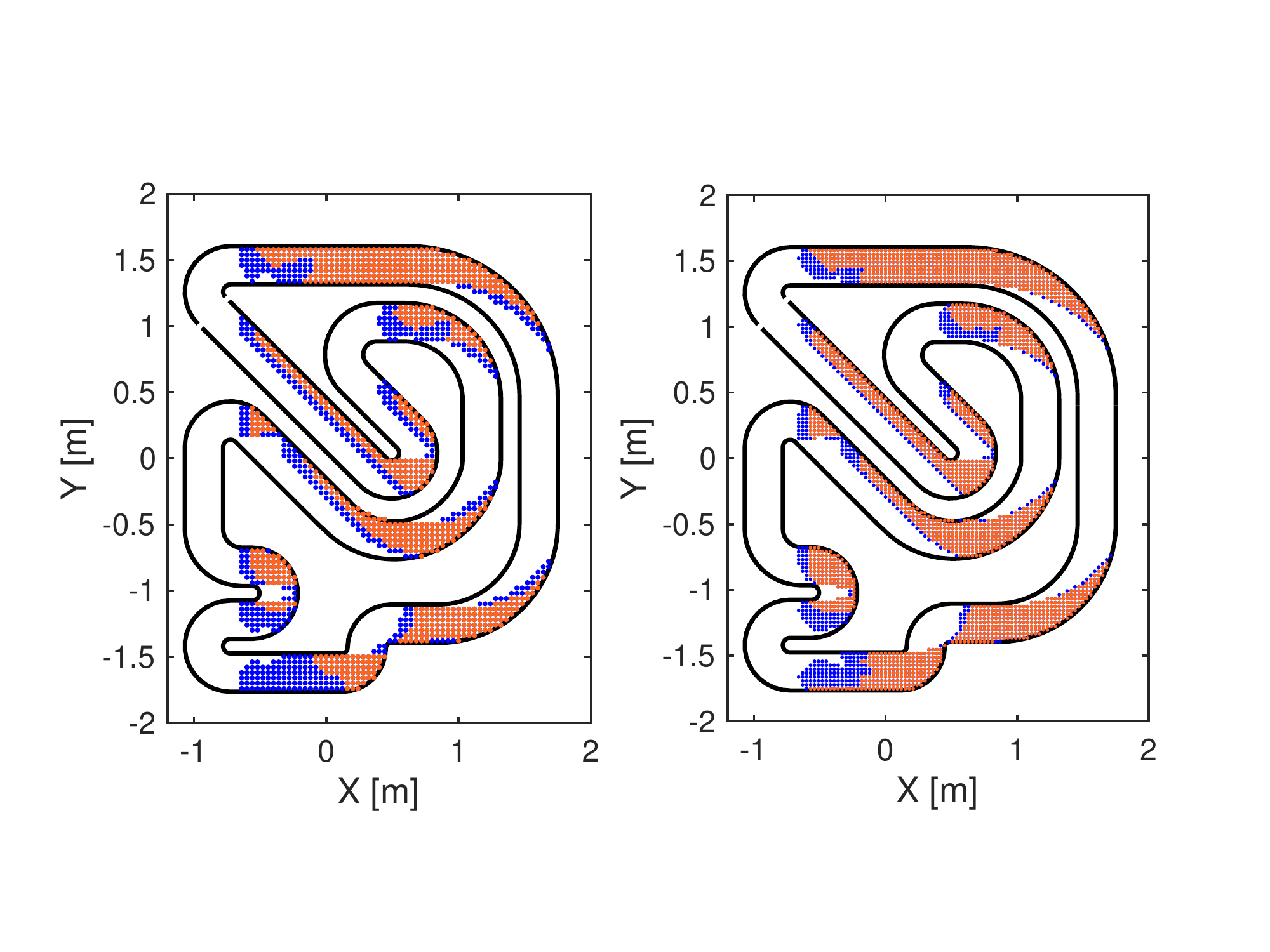}
  \caption{Red points indicate that the grid point is in both kernels, blue grid points are only in the viability kernel and not shown grid points are in neither kernel. Both plots show a slice through the kernels for the angle fixed to $\varphi = 180^\circ$ and a mode where the car is driving straight with 2 m/s. In the left figure 65'528'130 grid points are used in the calculation whereas the right we us a finer grid with 157'284'540 grid points.}\label{fig:ViabVsDiscTrack}
\end{figure}

Fig. \ref{fig:ViabVsDiscTrack}, visualizes the two kernels by comparing slices of the four dimensional kernels, this is achieved by fixing the angle $\varphi$ and the mode $q$, resulting in just the $X-Y$ plane which is easy to visualize. Fig. \ref{fig:ViabVsDiscTrack}, allows to see a qualitative difference between the viability and discriminating kernel for $r = 0.02$ and $r = 0.015$ (corresponding to 65'528'130 and 157'284'540 grid points respectively). Note that $r = 0.02$ implies that  the grid points are 4 cm apart; as a comparison the track is 37\,cm wide, and the size of the cars is about $10 \times 5$\,cm.
Fig. \ref{fig:ViabVsDiscTrack} shows that the difference between the two kernels also qualitatively becomes smaller if we use a finer grid. More precisely Fig. \ref{fig:ViabVsDiscTrack} shows all the viable points if the car is driving straight left with a forward velocity of 2\,m/s. For the illustrated angle and constant velocity, mainly the top straight and the bottom S-curve are of interest (depending on the driving direction). In this region the difference between the kernels and the effect of grid size is visible, for example, in the bottom S-curve, where the red region (which indicates that the grid point is in both kernels) gets significantly larger for the finer grid.

\section{Simulation results and sensitivity analysis} \label{sec:sim}
We start with a simulation study to investigate the effect of the various design choices. We use the full controller introduced in Section \ref{sec:HRHC}, which includes the viability-based path planner \eqref{eq:PredController} and the MPC tracking controller. The controls are applied to the continuous time bicycle model \eqref{eq:odeModel}, including control quantization motivated by the communication link in the real set-up, where both control inputs are represented using 8\,bit each.

The effect of the following tuning parameters in the path planner are examined:
\begin{itemize}
\item use of viability kernel, discriminating kernel, and no viability constraints;
\item grid size in $(X,Y,\varphi)$;
\item number of modes $N_m$;
\item path planning discretization time $T_{pp}$;
\item number of constant velocity segments $N_S$.
\end{itemize}

To measure a controller's performance, we consider the mean lap time, the number of constraint violations (i.e., the number of time steps the car is outside the track constraints) and the computation time required by the path planner. Each controller is simulated for 10'000 time steps; with a sampling time of 20\,ms, this corresponds to 200\,s or roughly 23 laps. The simulations are performed on a MacBook Pro with a 2.4 GHz Intel Core i7 processor, implemented in $\mathtt{C}$, using $\mathtt{gcc}$ as a compiler and $\mathtt{-O3}$ code optimization. Furthermore, we investigate the offline computation time and required memory for certain combinations, where the viability and discriminating kernel are computed in \emph{JULIA}, on a computer running Debian equipped with 16 GB of RAM and a 3.6 GHz Intel Xeno quad-core processor.
\subsection{Viability constraints and grid size}
\begin{table*}
\caption{Influence of the viability constraints}
\label{tab:ViabNonViab} 
\centering
\ra{1.3}
\begin{tabular}[t]{@{} l c c c c c c c c c c @{}}\toprule
\multicolumn{3}{c}{Controller} & & \multirow{2}{*}{
\parbox[b][{0.875cm}][b]{1.1cm}{mean lap time [s]}} & \multirow{2}{*}{
\parbox[c][{0.875cm}][b]{1.2cm}{$\#\,$constr. violations}} & \multicolumn{2}{c}{online comp. time [ms]} & \multirow{2}{*}{
\parbox[c][{0.875cm}][b]{1.3cm}{comp. time kernel [s]}} & \multicolumn{2}{c}{memory [MB]}\\
\cmidrule{1-3} \cmidrule{7-8} \cmidrule{10-11}
abbr. & kernel& grid &  &        &    & median & max    & &kernel & control \\
 \midrule
$\mathcal{C}^{nv}$ & No  & N/A &  & 8.77 & 10 & 43.71 & 334.23 & N/A & N/A & N/A  \\
$\bm{\mathcal{C}^{v}_c}$ & \textbf{ Viab } & \textbf{coarse}&    & \textbf{8.57} & \textbf{0} & \textbf{0.904} & \textbf{7.968} & \textbf{17957} & \textbf{65} & \textbf{1769 } \\
     $\mathcal{C}^{v}_f$ &  & fine &  &  8.39 & 3 & 0.870 & 7.557  & 43238 & 157 & 4246\\ 
$\mathcal{C}^{d}_c$ &Disc  & coarse  & & 8.60 & 1 & 0.870 & 7.533 & 156776 & 65 & 1769\\ 
      $ \mathcal{C}^{d}_f$& & fine & & 8.41 & 6 & 1.032 & 6.518  & 397674 & 157 & 4246\\
\bottomrule
\end{tabular}
\end{table*} 
We first compare our proposed controller that uses a path planner with viability constraints ($\mathcal{C}^{v}_f$, $\mathcal{C}^{d}_f$, $\mathcal{C}^{v}_c$ and $\mathcal{C}^{d}_c$), to a naive controller without viability constraints ($\mathcal{C}^{nv}$). Specially, we consider both the viability kernel and the discriminating kernel (superscript $v$ and $d$), and two different grid spacing ($r = 0.02$ and $r = 0.015$) referred to as coarse and fine (subscript $c$ and $f$). All the controllers use the parameters $N_m = 129$, $T_{pp} = 0.16$\,s and $N_S = 3$ in the path planner. $\mathcal{C}^{nv}$ is based on the implementation proposed in \cite{Liniger_2014} with the horizon increased from one to three and an improvement in the way constant velocities segments are concatenated in the path planner.

Table \ref{tab:ViabNonViab} compares all five controllers in terms of driving performance (mean lap time and constraint violations) as well as computation cost. We see that the number of constraint violation is negligible for all controllers as no controller triggers more than 10 violations (out of 10'000 steps). Focusing now on the mean lap time we can distinguish three groups: First, we see $\mathcal{C}^{nv}$ that has the highest mean lap time. Second are $\mathcal{C}^{v}_f$ and $\mathcal{C}^{d}_f$, which have very similar mean lap times, and are clearly the two fastest algorithms. Third, we have $\mathcal{C}^{v}_c$ and $\mathcal{C}^{d}_c$ whose mean lap times lie between the first two groups.

If we compare the online computation times of the controllers, then we can mainly differentiate between the controller without viability constraints ($\mathcal{C}^{nv}$) and controllers with viability constraints ($\mathcal{C}^{v}_f$, $\mathcal{C}^{d}_f$, $\mathcal{C}^{v}_c$ and $\mathcal{C}^{d}_c$) which are on average more than 40 times faster. As $\mathcal{C}^{nv}$ does not use any viability constraints to guide the path planning process, more branches of the tree have to be generated and checked, slowing down the computation, see Section \ref{subsec:limitations}. Furthermore, we see from Table \ref{tab:ViabNonViab} that the maximal computation time of $\mathcal{C}^{nv}$ is 334\,ms, which by far exceeds the sampling time of 20\,ms.

When comparing $\mathcal{C}^{nv}$ with the remaining controllers using the viability constraints, it is also important to investigate the offline computation time and the memory the viability lookup tables in \eqref{eq:PredController} require. First, for the controllers with viability constraints, we need to compute the corresponding kernels. Their computation time mainly depends on two factors: The choice between viability or discriminating kernel, and the used grid size. The number of grid points has a direct influence on the offline computation time, as the number of operations per grid point stays the same. The observation when comparing the chosen kernel algorithm is that the discriminating kernel algorithm is significantly slower. This has two reasons: first each iteration needs more operations than the viability kernel algorithm due to the disturbance \eqref{eq:discAlgo}; second the discriminating kernel algorithm needs more iterations to converge. For an implementation the required memory of the viability based path planner \eqref{eq:PredController} is also important, as \eqref{eq:PredController} requires to store a four dimensional lookup table for the kernel and a five dimensional lockup table for the viable inputs. The memory needed to store these lookup tables only depends on the grid size, and for the fine grid is 4.4 GB, and 1.8 GB for the coarse grid. Thus, we can conclude that, if the memory requirement and offline computation time are of no concern, then $\mathcal{C}^{v}_f$ and $\mathcal{C}^{d}_f$ should be preferred as they are the best performing controllers. However, if the available memory is limited, $\mathcal{C}^{v}_c$ and $\mathcal{C}^{d}_c$ with a coarse grid achieve good performance at a significantly lower memory requirement.

For the following discussions we will compare the other tuning parameters ($N_m$, $T_{pp}$ and $N_S$), of the path planner, while using $\mathcal{C}^{v}_c$ as the base comparison, marked as bold text in the tables.

\subsection{Number of modes, $N_m$} \label{sec:mode}
We examine the influence of $N_m$ by varying the number of modes, which depends on the grid imposed on the stationary velocity manifold, see Appendix \ref{app:N_m} for the different grids used. As expected, the lap time get slower the fewer modes are considered (mean lap time with $N_m = 89$ is 9.69\,s), but at the same time the computation time decrease by up to a factor of two. In all cases, the number of constraint violations is zero.

\subsection{Discretization time, $T_{pp}$}
The effect of time discretization was investigating by testing different values of $T_{pp} \in \{ 0.12,0.16,0.2,0.24\}$\,s. The results are summarized in Table \ref{tab:Tpp}, where we can see that larger $T_{pp}$ lead to lower computation times but the lap times getting slower. This effects come from the influence of $T_{pp}$ on the path planner model. As for the same mode (same constant velocity), a larger $T_{pp}$ implies that the car travels further. This may render this mode non-viable as the additional traveled distance may bring the car too close to the border. In contrast slower modes more likely stay inside the track. Thus, the path planner more likely chooses slow trajectories for larger $T_{pp}$, which leads to slower lap times. At the same time fewer trajectories are viable, which reduces the computation time. Thus, even though, a larger $T_{pp}$ implies a longer preview, the larger number of non-viable states mitigate this advantage. 
\begin{table}[h]
\caption{Influence of $T_{pp}$}
\label{tab:Tpp}
\centering 
\ra{1.3}
\begin{tabular}{@{}l c c c c @{}}\toprule
 & \multirow{2}{*}{
\parbox[b][{0.875cm}][b]{1.1cm}{mean lap time [s]}} & \multirow{2}{*}{
\parbox[c][{0.875cm}][b]{1.2cm}{$\#\,$constr. violations}} & \multicolumn{2}{c}{comp. time [ms]}\\
\cmidrule{4-5}
$T_{pp}$ &  &  & median & max \\
 \midrule
$0.12$    & 8.51 & 1 & 1.656 & 11.061  \\ 
\textbf{0.16}   & \textbf{8.57} & \textbf{0} & \textbf{0.904 }& \textbf{7.968} \\ 
$0.2$     & 8.79 & 1 & 0.720 & 6.169 \\ 
$0.24$    & 9.25 & 1  & 0.599 & 3.473 \\ 
\bottomrule
\end{tabular}
\end{table} 
\subsection{Number of constant velocity segments, $N_S$}
The influence of the number of constant velocity segments, or in other words the prediction horizon of the path planner is investigated by varying $N_S$ from two to four. In Table \ref{tab:N_S} we can see that more segments improve the performance but at the same time the computation time is heavily increased. The improved lap times came from the longer preview of the controller with larger $N_S$. The longer preview is achieved without changing the viability of a trajectory, which stands in contrast with larger $T_{pp}$. 
Furthermore, since both $T_{pp}$ and $N_S$ influence the preview window, we have observed that $N_S$ and $T_{pp}$ should be tuned simultaneously, e.g., for $N_S = 2$, $T_{pp} = 0.2$\,s achieves a faster mean lap time than $T_{pp} = 0.16$\,s.
\begin{table}
\caption{Influence of $N_S$}
\label{tab:N_S}
\centering 
\ra{1.3}
\begin{tabular}[h]{@{}l c c c c @{}}\toprule
 & \multirow{2}{*}{
\parbox[b][{0.875cm}][b]{1.1cm}{mean lap time [s]}} & \multirow{2}{*}{
\parbox[c][{0.875cm}][b]{1.2cm}{$\#\,$constr. violations}} & \multicolumn{2}{c}{comp. time [ms]}\\
\cmidrule{4-5}
$N_S$ &  &  & median & max \\
 \midrule
$2$   & 8.83 & 0 & 0.179 & 2.435 \\
\textbf{3 }  & \textbf{8.57} & \textbf{0} & \textbf{0.904} & \textbf{7.968} \\ 
$4$   & 8.56 & 1 & 17.015 & 114.278 \\ 
\bottomrule
\end{tabular}
\end{table} 
\subsection{Concluding remarks}
\begin{figure}[h]
\centering
\includegraphics[width = 0.65\textwidth]{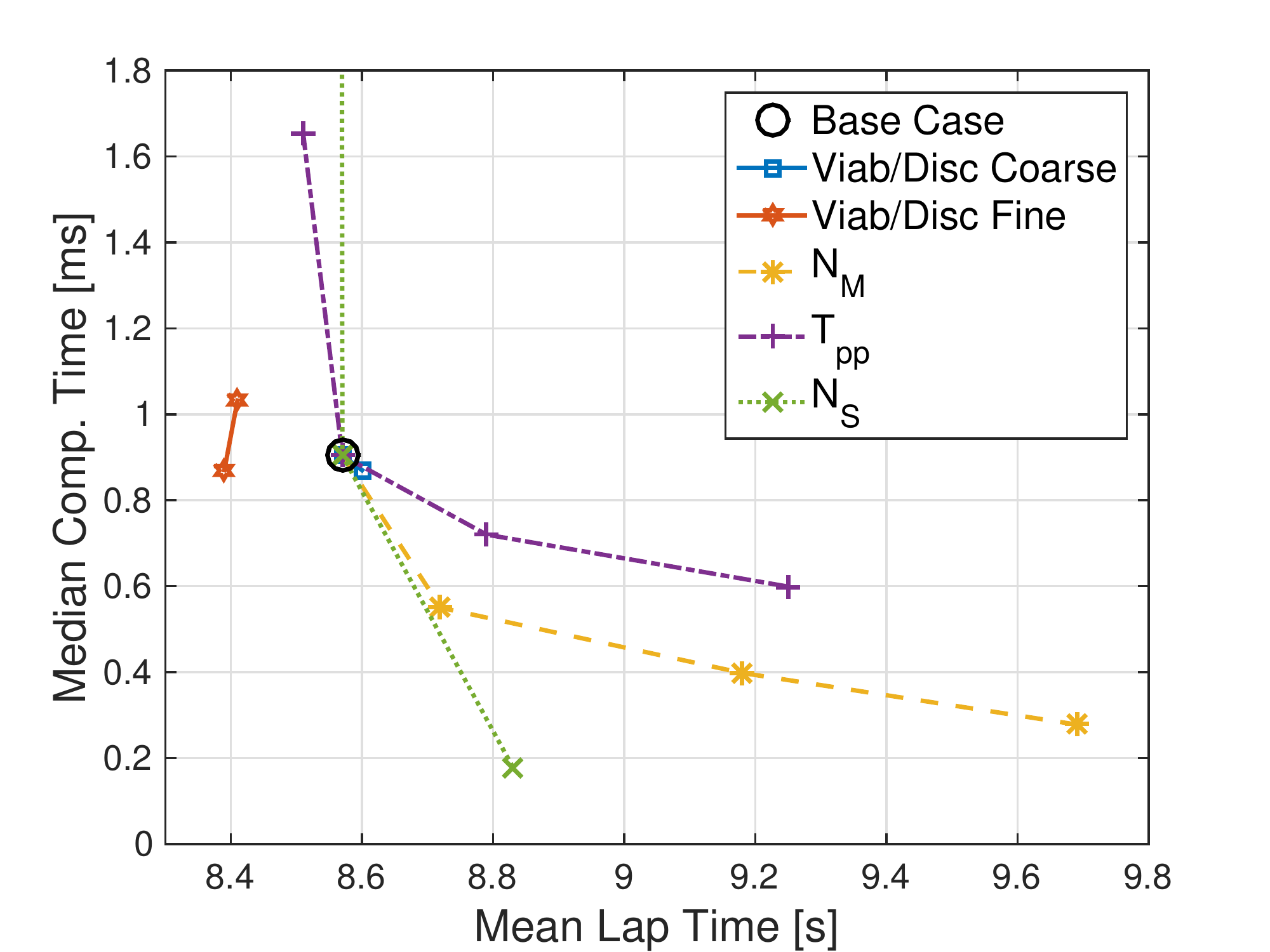}
  \caption{Visualization of the investigated cases, with the base case is marked with a circle. Solid lines Table \ref{tab:ViabNonViab}, dashed line Section \ref{sec:mode}, dash-doted line Table \ref{tab:Tpp} and doted line Table \ref{tab:N_S}. For visualization purposes the data for $\mathcal{C}^{nv}$ and $N_S = 4$ are excluded form the plot.}\label{fig:Pareto}
\end{figure}

The above discussion is summarized in Fig. \ref{fig:Pareto}, which allows us to draw the following conclusions: First, all the tested controllers based on kernel pruning outperform $\mathcal{C}^{nv}$, the state of the art controller based on \cite{Liniger_2014}, which for the horizon length considered does not even meet the real-time requirements. In contrast, the controllers with viability constraints and $N_S \leq 3$ are real-time feasible; put another way, the viability constraints allow one to use longer horizons in real-time than would be possible with the state of the art controller. Second, when comparing the variations of $\mathcal{C}^{v}_c$ with different values of $N_m$, $T_{pp}$ and $N_S$, we have observed that more modes $N_m$, shorter $T_{pp}$ and larger $N_S$ are beneficial for the performance. Furthermore, the numerical study indicates that, if a less powerful computer is used and the computation time has to be reduced, the most effective way to do so is to reduce $N_S$ to 2, see Fig. \ref{fig:Pareto}. Also note that $N_S = 4$ leads to no further improvement in the lap time, which indicates that $N_S = 3$ is a sufficiently long prediction horizon. This is also confirmed by using a terminal cost which captures the possible long term progress of a state, which improves the performance for $N_S = 2$, but does not help to reduce the lap time if $N_S \geq 3$.

Finally, we see from Table~\ref{tab:ViabNonViab} that the viability-based controller is marginally faster than the  discriminating-based controller. The latter, however, uses the proposed discriminating kernel approximation, which takes into account the discretization of the state space. Thus, the controller \eqref{eq:PredController} comes with additional viability guarantees. 

\subsection{Closed loop behavior of $\mathcal{C}^{v}_c$ and $\mathcal{C}^{d}_c$}

\begin{figure}[h]
\centering
\includegraphics[width = 0.65\textwidth]{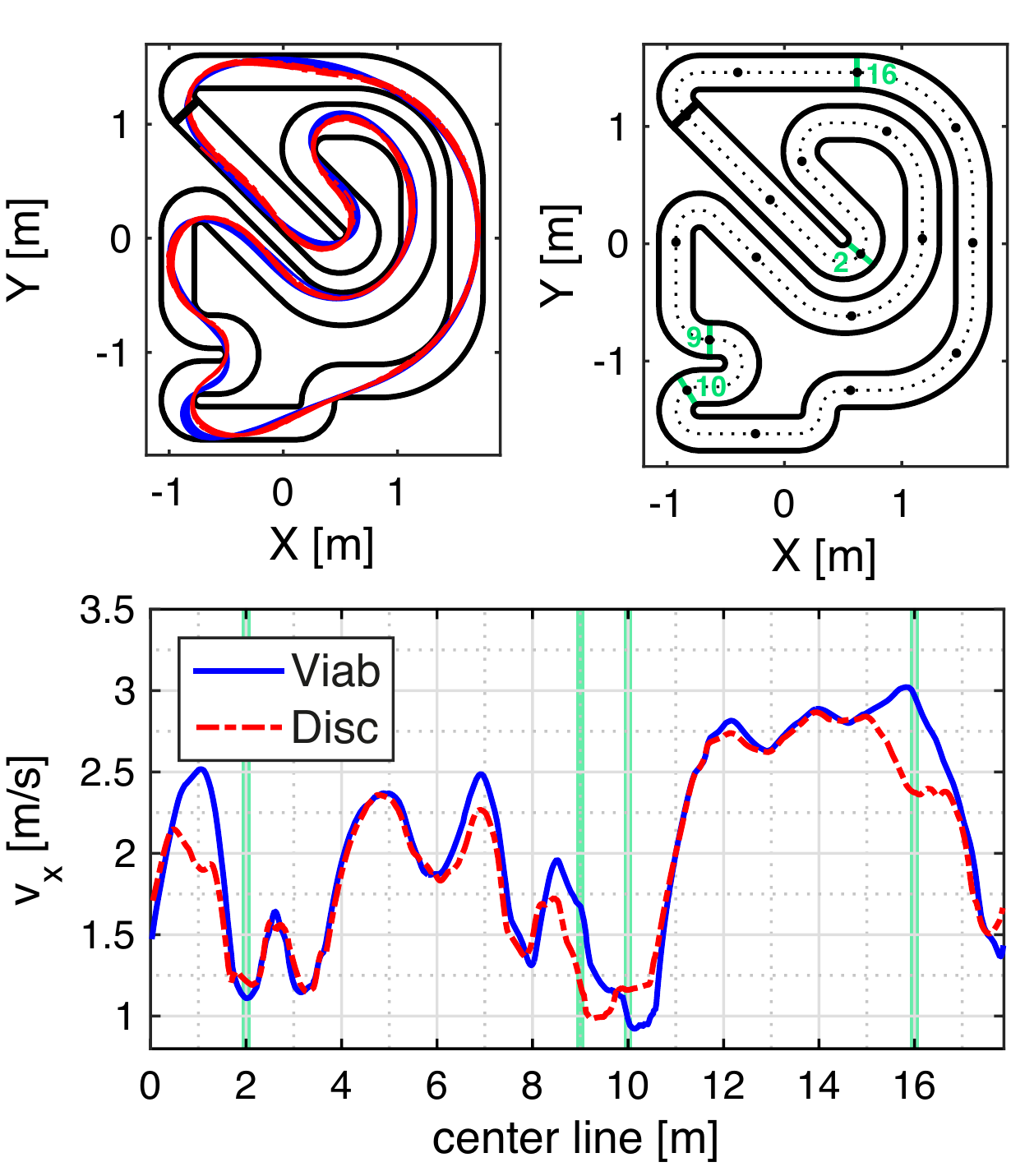}
  \caption{Top left: All trajectories of the controller using the viability kernel (in blue), and the discriminating kernel (in dash-doted red). Top right: To link relative position to the center line, every meter along the center line is marked with a point, whereas points of interest are marked with the length of the center line in green. Bottom: Velocity profile relative to the center line of one lap for both controllers, points of interest from the top right figure are marked with a green bar.}\label{fig:OneLap}
\end{figure}

Since the lap time of $\mathcal{C}^{v}_c$ and $\mathcal{C}^{d}_c$ are similar, it is interesting to investigate the driving behavior of the two controllers. Fig. \ref{fig:OneLap} shows one typical lap driven with $\mathcal{C}^{v}_c$ and $\mathcal{C}^{d}_c$. The beginning of the lap is in the top left corner marked with a line perpendicular to the track, and the cars race counter clockwise. The velocity is shown relative to center line, to make it possible to compare the velocity at a given point on the track; the projection on the center line is done as discussed in Section \ref{sec:HRHC}. We see that most of the track, the two controllers drive in a similar way and even have the same velocity. However, $\mathcal{C}^{d}_c$ drives somewhat slower coming into the curves, allowing higher velocity on the curve itself and higher exit velocity, the most extreme cases are marked with a green bars in Fig \ref{fig:OneLap}. It is interesting to see that, these two different driving styles lead to practically the same lap time.

\subsection{Obstacle avoidance}
\begin{figure}[h]
\centering
\includegraphics[width = 0.65\textwidth]{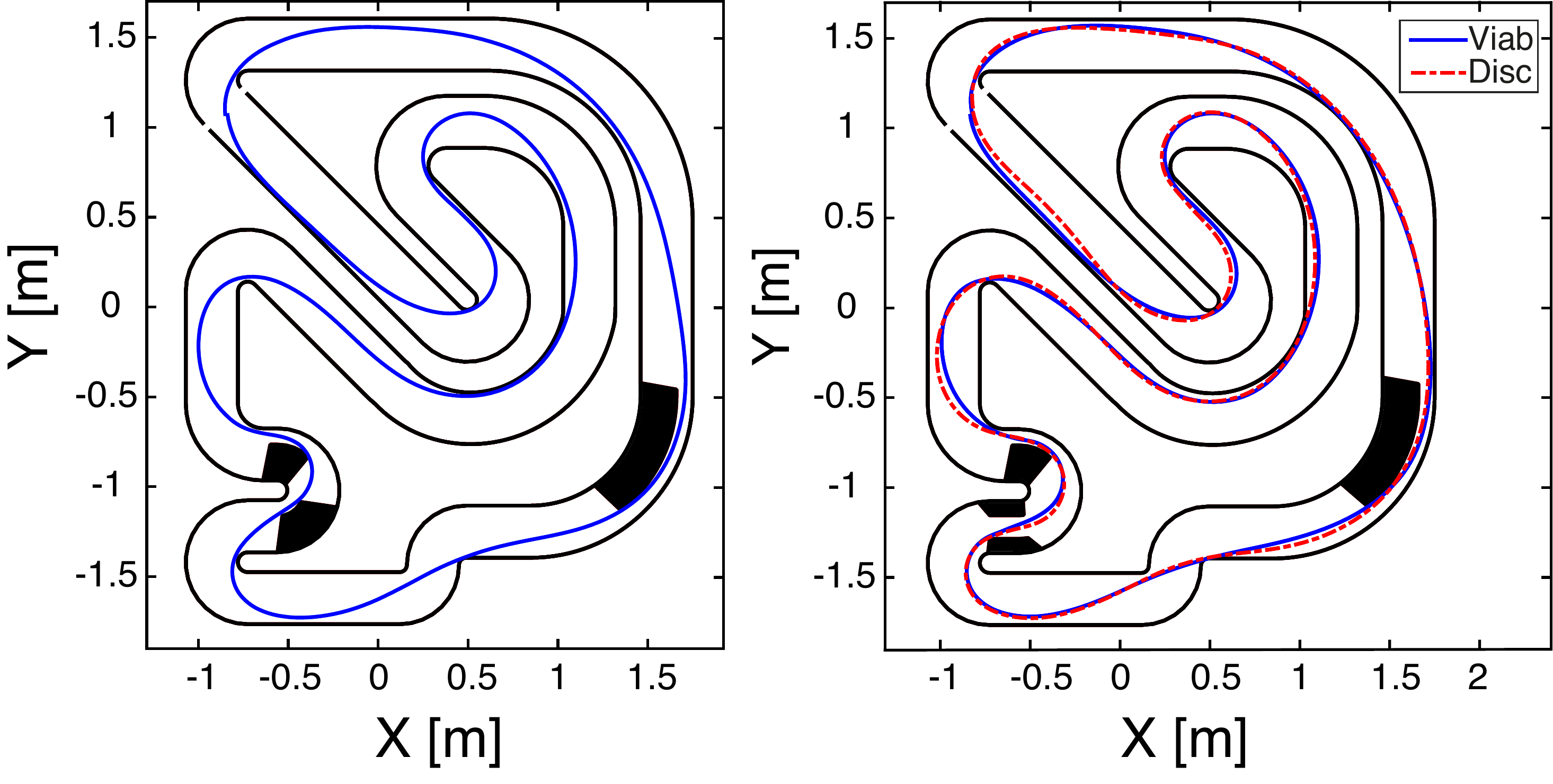}
  \caption{The best lap of the controllers for two obstacle constalation marked in black using the viability kernel (in blue), and the discriminating kernel (in dash-doted red). In the left obstacle configuration only the viability kernel based controller is able to navigate the course.}\label{fig:ObAvoid}
\end{figure}

To further highlight the difference between the two controllers $\mathcal{C}^{v}_c$ and $\mathcal{C}^{d}_c$ we included several obstacles at challenging position. The obstacle constraints can be included by modifying the constraint set $K$ and recomputing the viability and discriminating kernel. We tested two obstacle configurations shown in Fig. \ref{fig:ObAvoid}. For the first configuration (right plot in Fig. \ref{fig:ObAvoid}) $\mathcal{C}^{v}_c$ successfully avoids the obstacles, but the discriminating kernel collapses to the empty set. The collapse is due to the conservatism added by considering the space discretization  and the fact that the model cannot stop. The second obstacle configuration is easier to navigate and both controllers are able to find paths around the obstacles. However, $\mathcal{C}^{v}_c$ is significantly faster than $\mathcal{C}^{d}_c$ with a mean lap time of 9.066 compared to 9.272\,s.

\section{Experimental results} \label{sec:exp}
To verify the simulation results, we implemented $\mathcal{C}^{v}_c$ and $\mathcal{C}^{d}_c$ in an experimental set-up. The set-up consists of 1:43 Kyosho dnano RC cars, which are driven autonomously around the race track shown in Fig. \ref{fig:track}. The control signals are sent to the cars by an external computer via Bluetooth and an infrared-based vision system captures the cars’ current position, orientation and velocity. We refer the interested reader to \cite{Liniger_2014} for a more detailed description of the hardware set-up. The controllers were implemented on a desktop computer running Ubuntu 14.04 OS, equipped with 4 GB of RAM and a 3.5 GHz Intel i7 quad-core processor. The tracking MPC problem is solved using FORCES Pro \cite{FORCESPro}.

\begin{figure}[h]
\centering
\includegraphics[width = 0.65\textwidth]{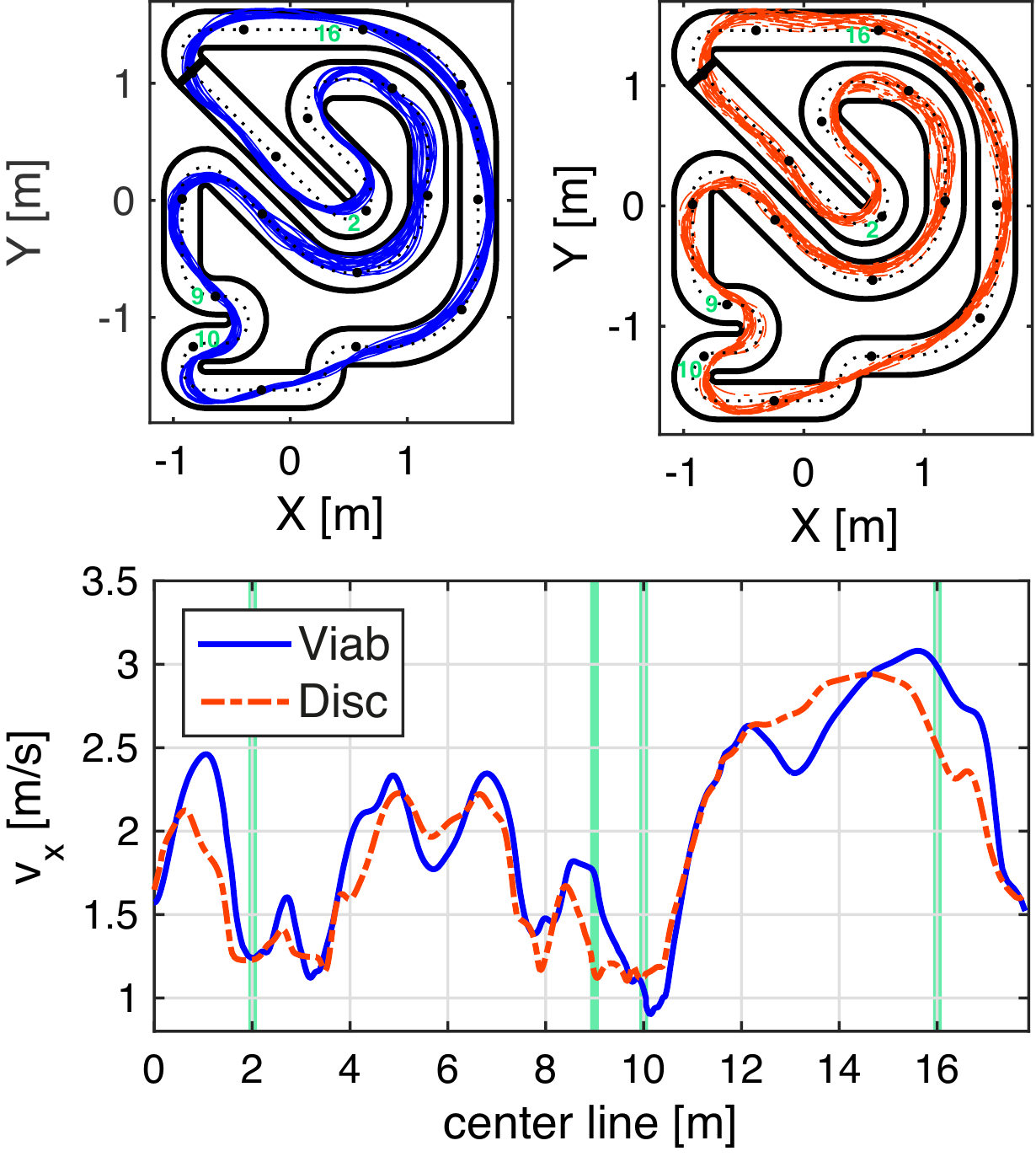}
  \caption{All laps of $\mathcal{C}^{v}_c$ (in blue) and $\mathcal{C}^{d}_c$ (in dash-doted red) as well as the velocity profile of one lap.}\label{fig:Exp}
\end{figure}

For the experiments we implemented the two controllers $\mathcal{C}^{v}_c$ and $\mathcal{C}^{d}_c$ due to their good performance, and because they are feasible in terms of memory (compared to $\mathcal{C}^{v}_f$ and $\mathcal{C}^{d}_f$). We ran the experiment for 200\,s and extracted all completed laps, shown in Fig. \ref{fig:Exp}, together with the velocity profile of one lap. We see from Table \ref{tab:exp} that constraint violations occur more frequently than in simulation. This is mainly due to a significantly larger model mismatch when using the real car compared to the bicycle model used in simulation. The increased model mismatch can also been seen by the increased spread of the trajectories between the simulation and the experimental results, compare Fig. \ref{fig:OneLap} and \ref{fig:Exp}.

One way to deal with the model mismatch is to tighten the track constraints of the controller to illustrate this we impose track constraints that are  1.5\,cm away from the track boundary in the viability computations, and 0.5\,cm in the MPC; for comparison the miniature cars are $10 \times 5$\,cm in size. This provides the controller a certain margin of errors but of course comes at the cost of a potential increase in lap time. Constraint violation, are counted when the car is closer than 0.5\,cm to the track boundary. The model mismatch can also lead to infeasibilities in the controller. This is, for example the case for the path planner when the current state is not inside the viability kernel. In such a case we resolve infeasibilities in the path planning process using the following heuristics: First we find a viable neighboring cell and solve the path planning problem pretending to be in the closest viable neighboring grid cell. If none of the neighboring grid cells is viable, then an emergency stop is initiated, and the controller is restarted if the car is back in a viable state. Infeasibilities in the lower level MPC problem are dealt with by using soft constraints \cite{Kerrigan2000}.

We see from Fig. \ref{fig:Exp} that $\mathcal{C}^{d}_c$ drives more conservatively, a feature we have also seen in the simulation studies before. By breaking earlier, $\mathcal{C}^{d}_c$ is able to achieve higher velocities on the curves itself and higher exit velocities. In contrast to the simulation, however, the difference in the mean lap times between $\mathcal{C}^{d}_c$ and $\mathcal{C}^{v}_c$ is larger, see Table \ref{tab:exp}, though the best and the worst laps are close with lap times in the range 8.66 to 9.54\,s for $\mathcal{C}^{v}_c$ and 8.64 to 9.44\,s for $\mathcal{C}^{d}_c$. Also the number of constraint violations, is nearly identical. For a video comparison of the two controllers, see \url{https://youtu.be/RlZdMojOni0}.
\begin{table}[h]
\caption{Experimental implementation of $\mathcal{C}^{v}_c$ and $\mathcal{C}^{d}_c$. }
\label{tab:exp}
\centering 
\ra{1.3}
\begin{tabular}{@{}l c c c c @{}}\toprule
 & \multirow{2}{*}{
\parbox[b][{0.875cm}][b]{1.25cm}{median lap time [s]}} & \multirow{2}{*}{
\parbox[c][{0.875cm}][b]{1.2cm}{$\#\,$constr. violations}} & \multicolumn{2}{c}{comp. time [ms]}\\ 
\cmidrule{4-5}
Kernel &  &  & median & max \\
 \midrule
Viab   &  8.86 & 42 & 1.125 & 12.936 \\
Disc    & 8.94 &  46 & 1.176 & 11.055 \\
\bottomrule
\end{tabular}
\end{table} 

\section{Conclusion} \label{sec:conclusion}

In this work we showed that the existing hierarchical controller of \cite{Liniger_2014} can be improved by incorporating the viability kernel in the path planning phase. As a result, the path planner only generates viable trajectories that are recursively feasible, while reducing the computation time. This in turn allows the use of longer prediction horizons, which generally leads to better performance in terms of lap times and constraint violations.

To compensate for discretization errors in the computation of the viability kernel, we formulated the viability computation problem as a game between the uncertain initial condition introduced by the state discretization and the control input. The resulting kernel was calculated using the discriminating kernel algorithm, which allows us to derive an inner approximation of the normal viability kernel. Furthermore, the new kernel guarantees that there exists a control input that keeps the system within the kernel even if the current state is not on a grid point but somewhere within the cell around a grid point. 

In a numerical study we investigated the influence of different parameters on the performance of the controller using the viability conststraints in the path planning phase of our controller, and compared the performance if the standard viability kernel is used compared with our inner approximation. Although the closed loop behavior of the controller using the standard viability kernel and the proposed discriminating kernel approximation are different, the performance in terms of lap time of the two controllers is very similar. The controller based on the proposed discriminating kernel approximation seems to drive with more foresight and is less aggressive. The same behavior was also observed in our experimental implementation.

This work has mainly focused on the path planning step in the hierarchical controller, whereas the lower level MPC is kept simple. In future work we will investigate how the lower level can be modified to improve the driving performance. For example by considering the model uncertainty in the MPC design, as proposed in \cite{Carrau2016}. Furthermore, we are investigating the use of the path planning model and the viability constraints in a racing game with multiple opposing cars.

\section*{Acknowledgment}
The authors would like to thank Xiaojing Zhang for the helpful discussions and his advice.

\begin{appendices} 
\section{Finite inner approximation} \label{app:InnerApprox}
Recall from \eqref{eq:finiteGuarantee} that the finite viability kernel is an inner approximation of the discrete viability kernel with respect to an extended set-valued map. For the discriminating kernel the authors are not aware of  similar results. Even though the basic algorithms \eqref{eq:viabAlgo} and \eqref{eq:discAlgo} are similar, the discretization of the disturbance space $V_h \subset V$ required in the discriminating kernel algorithm ``weakens" the uncertainty. This may lead to points in the \emph{finite discrete} discriminating kernel $\Disc_{G^r_h}(K_h)$ which are outside the \emph{discrete} discriminating kernel $\Disc_{G^r}(K) \cap X_h$.

In the following we present a method for addressing this problem for our dynamics \eqref{eq:dynSysOfInterest}. This is achieved by first showing that by only discretizing space, but not the disturbance space, the ``semi-finite" discriminating kernel is an inner approximation of $\Disc_{G^r}(K) \cap X_h$. And second by showing that for every discrete disturbance $v_h \in V_h$ there exists a set of disturbances $\tilde{V} \subset V$ which does not change the finite dynamical system $x_{h,k+1} \in G(x_{h,k},v_{h,k})$. This inside then allows to formulate a modified discriminating kernel algorithm, which guarantees that $\Disc_{G^r_h}(K_h) \subset \Disc_{G^r}(K) \cap X_h$.
\subsection{Continuous Disturbance Space}
Let us first look into the ``semi-finite" discriminating kernel where the state space is discretized but the disturbance input is continuous. To this end let us introduce a new set-valued map $\tilde{G}^r_h:X_h\times V \rightsquigarrow X_h$, $\tilde{G}^r_h(x_h,v) := G^r(x_h,v) \cap X_h$, which leads to the following finite dynamical system, 
\begin{align} \label{eq:semiFinitDynSys}
x_{h,k+1} \in \tilde{G}^r_h(x_{h,k},v_k)\,,
\end{align}
where $v_k \in V$. The following result holds for the new ``semi-finite" discriminating kernel $\Disc_{\tilde{G}_h^{r}}(K_h)$
\begin{proposition} \label{prop:innerApproxDisc}
Let $G:\mathbb{R}^n\times V \rightsquigarrow \mathbb{R}^n$ be an upper-semicontinuous set-valued map with compact values, $K$ be a closed subset of $\textnormal{Dom}(G)$, and $V$ a compact set. Let $r$ be such that, $\forall v \in V\, \forall x \in \textnormal{Dom}(G^r) \cap X_h\,, \; G^r(x,v) \cap X_h \neq \emptyset$. Then, $\Disc_{\tilde{G}_h^{r}}(K_h) \subset \Disc_{G^r}(K) \cap X_h$.
\end{proposition}
\begin{IEEEproof}
The proof is an extension of the proof of \emph{Proposition 4.1} in \cite{saintPierre94} to the case of the discriminating kernel. We have that for any $v\in V$, $\tilde{G}^r_h(x_h,v) \subset G^r(x_h,v)$ and additionally $K_h \subset K$. By the definition of the discriminating kernel, we know that for all $x_h \in \Disc_{\tilde{G}_h^{r}}(K_h)$ there exists a solution to the finite dynamical system \eqref{eq:semiFinitDynSys} starting from this grid point, which forever stays within $K_h$, for any sequence of disturbance inputs $v_k$. As $ \tilde{G}^r_h(x_h,v) \subset G^r(x_h,v)$ and $K_h \subset K$, there also exists a solution to $x_{k+1} \in G^r(x_k,v_k)$, starting at the same grid point which stays in $K$, for any disturbance input sequence. One trivial solution to $x_{k+1} \in G^r(x_k,v_k)$, which fulfills the property is same as the one of the finite dynamical system. Therefore, if $x_h \in \Disc_{\tilde{G}_h^{r}}(K_h)$ the grid point is also an element of $\Disc_{{G}^{r}}(K)$, which concludes the proof. 
\end{IEEEproof}
Proposition \ref{prop:innerApproxDisc} would allow us to state the discrete space counterpart of Proposition \ref{prop:PropertiesAlgo}. However, $\Disc_{\tilde{G}_h^{r}}(K_h)$ can not be computed, thus we proceed by establishing a link between $\tilde{G}^r_h(x_h,v)$ and ${G}^r_h(x_h,v_h)$, which allows to use Proposition \ref{prop:innerApproxDisc} in the ``fully finite" case.

\subsection{Discrete Disturbance Space}
\begin{figure}[h]
\centering
\includegraphics[width = 0.65\textwidth]{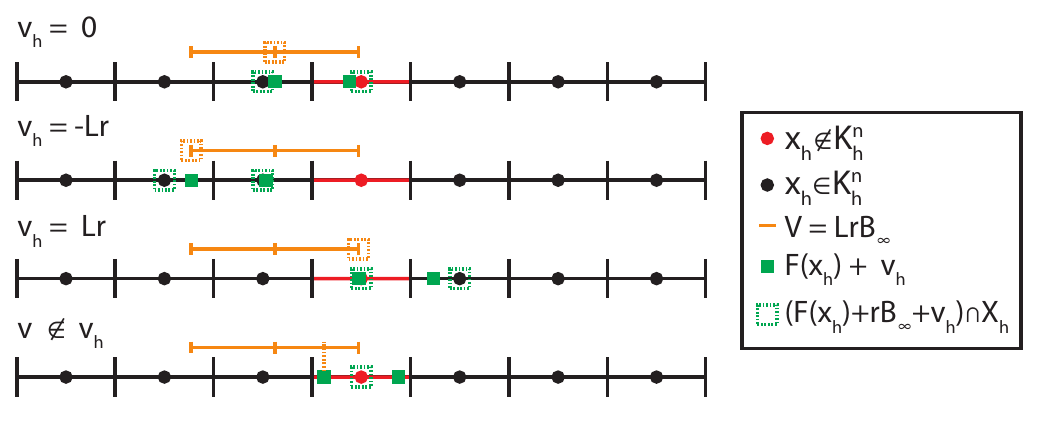}
  \caption{Visualization of an instance where the fully finite algorithm gives a wrong answer for a given grid point. In the example a simple 1-D system $F(x)$ with two discrete values is considered. For all finite disturbances $((F(x_h) + rB_\infty + v_h)\cap X_h) \cap K_h^n$ is non empty, as the first the lines illustrate, however, there exists a $v \notin v_h$, where the condition is not fulfilled.}\label{fig:ProblemFiniteDist}
\end{figure}

By discretizing the disturbance space $V_h \subset V$, the possible actions of the disturbance input are reduced, which can result in an wrong classification of states to lay within the discriminating kernel. More precisely we can only guarantee that there exists a trajectory to the difference inclusion which stays in $K_h$ for all sequences $v_{h,k}$. However, this does not necessary hold for all continuous disturbance sequences $v_k$, as illustrated in Fig. \ref{fig:ProblemFiniteDist}. Even under restrictive assumptions on the used grids, e.g., square grids, and fine disturbance grid this problem is not solved.

However, it is possible to establish a link between the continuous and discrete disturbance, for the dynamical system used in Proposition \ref{prop:PropertiesAlgo}, $x_{k+1} \in F(x_k) + v_k$ with $v_k \in V = LrB_\infty$, where the uncertainty enters additive and the disturbance space is a box. We begin with stating the assumptions on the state and disturbance grid.

\begin{assumption} \label{as:grid} $ $
\begin{itemize}
\item $X_h \subset X$ is a square grid with a spacing of $2r$ in the infinity norm.
\item $V_h \subseteq V$ is a square grid, where any point $v \in V \subset \mathbb{R}^d$ has $2^d$ neighboring points in $V_h$ which are closer than $2r$ in the infinity norm.
\end{itemize}
\end{assumption}
The assumption on the state grid is necessary to exclude special cases that can arise with non square grids. The assumption on the disturbance grid generates a grid where $V = LrB_\infty$ is gridded with $\lceil L \rceil + 1$ grid points in each direction, including the corner points and grid points at the boundary of the set. Additionally the disturbance grid spacing is smaller than the state grid spacing. Therefore, it is impossible that two neighboring disturbance grid points can move $F(x_h)$ further than to the next grid cell.

In addition to Assumption \ref{as:grid} we need an assumption on the set of allowed controls. As the proposed method will check every element of $F(x)$, it is necessary that $F(x)$ is a finite set, which is the case if the set of admissible controls is finite. To achieve this, the set of control inputs $U$ is discretized, leading to the following finite subset $U_h \subseteq U$. Notice that $U_h$ does not need any additional structure. 

\begin{corollary} \label{co:proj}
If $X_h$ and $V_h$ fulfill Assumption \ref{as:grid}, then for all $x_h \in X_h$, $v_h \in V_h$, and $u_h \in U_h $, it holds that
\begin{align*}
&x^*_h = (f(x_h,u_h)+v_h+rB_\infty) \cap X_h\,, \nonumber \\
\Leftrightarrow &f(x_h,u_h)+v_h \in x^*_h+rB_\infty\,.
\end{align*}
\end{corollary}
\begin{IEEEproof}
It is easy to see that the two statements are equivalent, as in the first statement the point is projected onto a grid point whenever the it is closer than $r$ to this grid point in the infinity norm, and the second statement states that the point is in a cell around the grid point with radius $r$ in the infinity norm. Both statements are only true if $\| x^*_h - (f(x_h,u_h)+v_h) \|_\infty \leq r$
\end{IEEEproof}

\begin{figure}[h]
\centering
\includegraphics[width = 0.65\textwidth]{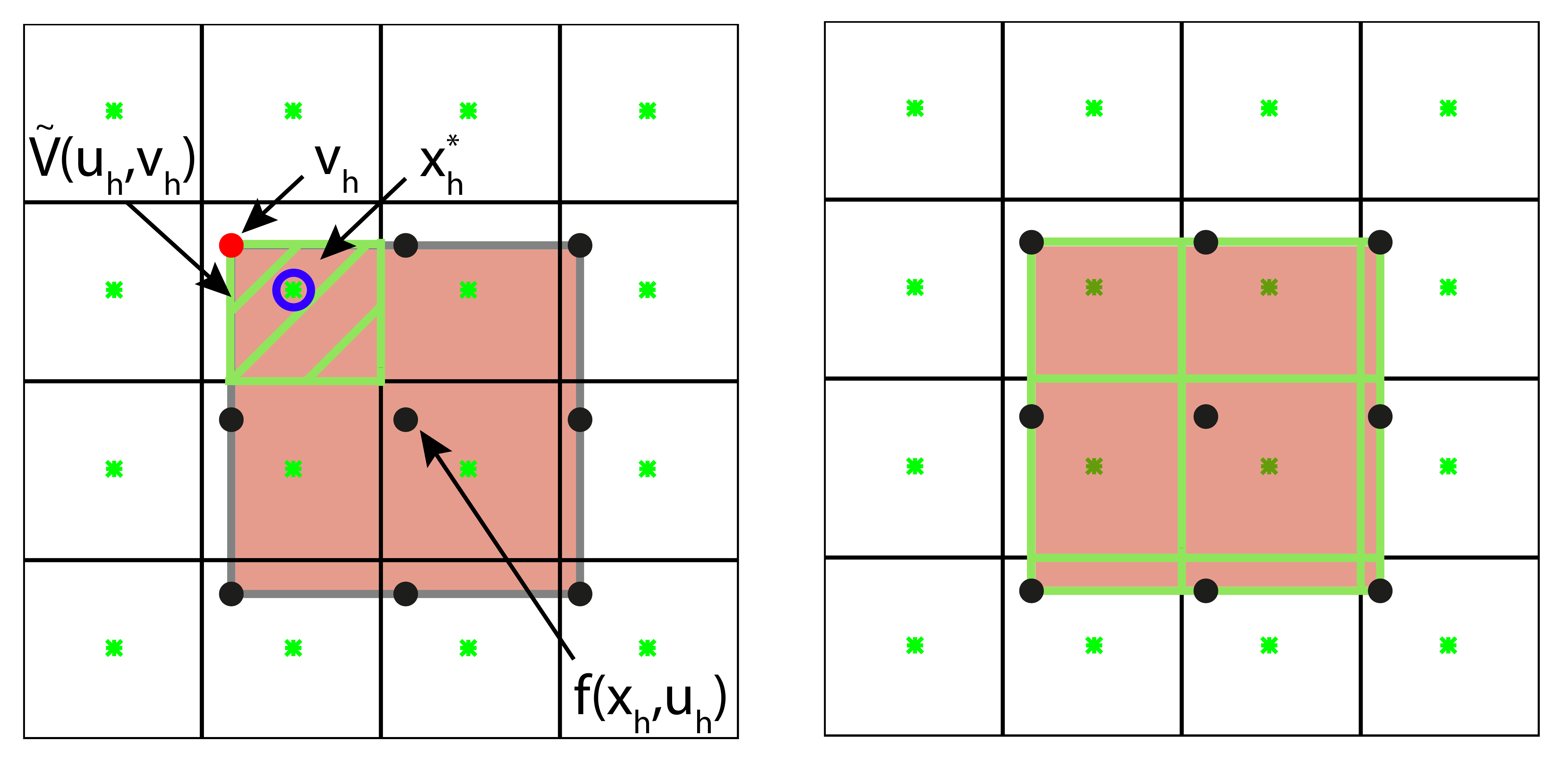}
  \caption{Visualization of the box $\tilde{V}(u_h,v_h)$, on the left for one disturbance grid point, and on the right all resulting $\tilde{V}(u_h,v_h)$ boxes for one control input are shown as green boxes.}\label{fig:Vhat}
\end{figure}

From Corollary \ref{co:proj}, we can see that there exists a neighborhood around $v_h$, which still leads to the same $x^*_h$. Thus, for a given $x_h$, $u_h$ and $v_h$, we can compute the subset of all continuous disturbances $\tilde{V}(x_h,u_h,v_h) \subseteq V$ which maps to the same state grid point $x_h^*= (f(x_h,u_h) + v_h + rB_\infty)\cap X_h$,
\begin{align*}
\tilde{V}(x_h,u_h,v_h) =  \{v \in V | \nonumber  \|x_h^* - (f(x_h,u_h) + v)\|_\infty \leq r   \}\,,
\end{align*}
see Fig. \ref{fig:Vhat} for an illustration of $\tilde{V}(x_h,u_h,v_h)$. In the interest of readability the dependency of $\tilde{V}$ on the grid point $x_h$ will subsequently be left out, and we just refer to $\tilde{V}(u_h,v_h)$.

Thus, by only looking at finite disturbance $v_h$, but considering $\tilde{V}(u_h,v_h)$ it is possible to make a statement about the continuous disturbance. In the following we will discuss how the set $\tilde{V}(u_h,v_h)$ can be used to formulate a modified discriminating kernel algorithm which is equivalent to the ``semi-finite" case described in Proposition \ref{prop:innerApproxDisc}.

Given an instance of the discriminating kernel algorithm with a grid point $x_h \in K_h^n$, we can compute the set of all $u_h,\;v_h$ such that the intersection of $(f(x_h,u_h) + v_h + rB_\infty)\cap X_h$ and $K_h^n$ is non-empty,
\begin{align} \label{eq:I_uv}
I_{u,v}(x_h) &= \{u_h\in U_h , \; v_h \in V_h | \; \nonumber \\
&(f(x_h,u_h) + v_h +rB_\infty)\cap X_h \in K_h^n \}\,.
\end{align}
\begin{proposition} \label{prop:Union}
Given $x_h \in K_h^n$ and $I_{u,v}(x_h)$,  it holds that for all $v \in V$ there exists a $u_h \in U_h$ such that
\begin{align*}
(f(x_h,u_h) + v + rB_\infty)\cap X_h \in K^n_h \,,\nonumber
\end{align*}
if and only if, $\bigcup_{I_{u,v}(x_h)} \tilde{V}(u_h,v_h) = V$
\end{proposition}
\begin{IEEEproof}
The \emph{if} part directly follows from the definition, as all continuous disturbances $v \in V$ are considered. The \emph{only if} part is satisfied by Assumption \ref{as:grid}, which ensures that for a control $u_h$ the union of all $\tilde{V}$ is equal to $V$.
\end{IEEEproof}
Similar to the proof of Proposition \ref{prop:PropertiesAlgo}, by using \cite{Cardaliaguet1994}[Equation (4)], we know that for the ``semi-finite" discriminating kernel it holds that if $x_h \in \Disc_{\tilde{G}_h^{r}}(K_h)$, for all $v\in V$ there exists a $u_h \in U_h$ such that $(f(x_h,u_h) + v + rB_\infty)\cap X_h \in \Disc_{\tilde{G}_h^{r}}(K_h)$. Thus, $\bigcup_{I_{u,v}(x_h)} \tilde{V}(u_h,v_h) = V$ allows us to compute $\Disc_{\tilde{G}_h^{r}}(K_h)$ by only considering a finite number of disturbances. However, we need to compute $\bigcup_{I_{u,v}(x_h)} \tilde{V}(u_h,v_h) = V$ which may be hard. Thus, we now focus on calculating and approximating of $\bigcup_{I_{u,v}(x_h)} \tilde{V}(u_h,v_h) = V$. Notice first that the set $\tilde{V}(u_h,v_h)$ is always a box. Therefore, they can be parameterized as
\begin{align*}
\tilde{V}(u_h,v_h) = \{ v \in V | \underline{v}_j \leq v_j \leq  \overline{v}_j, \forall j = 1,...,d  \}\,.
\end{align*}
Where, the subscript $j$ refers to each dimension of the uncertainty $V \subset \mathbb{R}^d$ and for each of these dimensions the lower and upper bound can be calculated by,
\begin{subequations}\label{eq:Vbounds}
\begin{align*} 
&\underline{v}_j(u_h,v_h)  =    \min(x_{h,j}^* - f(x_h,u_h)_j -  r ,-Lr)\,,\\
&\overline{v}_j(u_h,v_h)  =    \max(x_{h,j}^* - f(x_h,u_h)_j +  r , Lr)\,.
\end{align*}
\end{subequations}
As the union of boxes is in general not convex, checking that $\bigcup_{I_{u,v}(x_h)} \tilde{V}(u_h,v_h) = V$ is a combinatorial problem. However, by approximating the union conservatively we are able reduce the computational burden while guaranteeing an inner approximation.

To approximate the union notice that $v_h \in \tilde{V}(u_h,v_h)$ always holds. Thus we first find a box approximation for the union of the boxes around each $v_h$ grid point, and then calculate the union of the resulting $|V_h|$ boxes, which can be done by checking if neighboring boxes overlap. To compute a box approximation of the union at one disturbance grid point let us define $I_{u}(x_h,v_h) = \{u_h\in U_h  | \; (f(x_h,u_h) + v_h +rB_\infty)\cap X_h \in K_h^n \}$, which is a set similar to $I_{u,v}(x_h)$. Here we propose two approximations of $\bigcup_{I_{u}(x_h,v_h)} \tilde{V}(u_h,v_h)$, first the maximal volume box $\max_{I_{u}(x_h,v_h)} \text{Vol}(\tilde{V}(u_h,v_h))$ and second the intersection of all boxes $\bigcap_{I_{u}(x_h,v_h)} \tilde{V}(u_h,v_h)$. Both approximations result in a box and are an inner approximation of the union.

Thus, we propose the modified discriminating kernel algorithm outlined in Algorithm \ref{algo:innerApproxDisc}. The algorithm guarantees an inner approximation of the discriminating kernel in the following sense, $\Disc_{{G}_h^{r}}(K_h) \subset\Disc_{\tilde{G}_h^{r}}(K_h) \subset\Disc_{{G}^{r}}(K) \cap X_h$, the first inner approximation follows due to the approximation of $\bigcup_{I_{u}(x_h,v_h)} \tilde{V}(u_h,v_h)$ and the second due to Proposition \ref{prop:innerApproxDisc}.

\begin{algorithm}[h!] \label{algo:innerApproxDisc}
	initialization $K^0_h = K_h$ and $n = -1$\;
    \Do{$K^n_h \neq K^{n+1}_h$}{
		$n = n+1$\;
		\For{all $x_h \in K^n_h$}{
			calculate $I_{u,v}(x_h)$, in \eqref{eq:I_uv}\;
			calculate the corresponding $\tilde{V}(u_h,v_h)$\;
			$\forall v_h \in V_h$ compute $\max_{I_{u}(x_h,v_h)} \text{Vol}(\tilde{V}(u_h,v_h))$\;
			check if neighboring boxes overlap\;
			\If{Yes}{
				$x_h \in K^{n+1}_h$
			}
		}
    }
    \KwResult{ $\Disc_{G^r_h}(K_h) = K^n_h$} 
  \caption{Modified Discriminating Kernel Algorithm}
\end{algorithm}

Note that the union approximation is not necessary if one control input is robust with respect to all discrete disturbances. Because by virtue of Assumption \ref{as:grid}, we know that in this case $\bigcup_{I_{u,v}(x_h)} \tilde{V}(u_h,v_h) = V$.

As the proposed algorithm constructs an inner approximation in the case of a finite disturbance, it is possible to state the following finite version of Proposition \eqref{prop:PropertiesAlgo}.

\begin{proposition} \label{prop:controlInnerApproxAppendix}
Consider the finite dynamical system corresponding to the extended and discretized system of \eqref{eq:dynSysOfInterest},
\begin{align*}
x_{h,k+1} \in G^r_h(x_{h,k}) =  (F(x_{h,k}) + v_h + r B_\infty)\cap X_h\,,
\end{align*}
where $v_h \in V_h$ is the discretization of $L r B_\infty$ according to Assumption \ref{as:grid}. And if $\Disc_{G^r_h}(K_h)$ is computed with the Algorithm \ref{algo:innerApproxDisc} and Assumption \ref{as:grid} holds, then the following properties hold:
\begin{itemize}
\item[1)] $\Disc_{G^r_h}(K_h)$ is a viability domain of $F^r_h$.
\item[2)] For all $x_h \in \Disc_{{G}^r_h}(K_h)$ and for all $\hat{x} \in x_h + rB_\infty$, there exists a $u_h \in U_h$ such that $f(\hat{x},u_h) \in \Disc_{{G}^r_h}(K_h)+rB_\infty$.
\end{itemize}
\end{proposition}
\begin{IEEEproof} 
1) Identical to the proof of Proposition \ref{prop:PropertiesAlgo} statement 2), as $\Disc_{G^r_h}(K_h)$ is a discriminating domain and $v=0$ is an allowed disturbance.\newline
2) By using the finite version of \cite{Cardaliaguet1994}[Equation (4)], and the guarantee of the inner approximation $\Disc_{{G}_h^{r}}(K_h) \subset\Disc_{{G}^{r}}(K) \cap X_h$ we know that for all $v \in LrB_\infty$ there exists a $u_h \in U_h$ such that $(f(x_h,u_h) + v + rB_\infty)\cap X_h \in \Disc_{{G}_h^{r}}(K_h)$. Furthermore, by Corollary \ref{co:proj} we can reformulate the discriminating kernel condition as, for all $v \in LrB_\infty$ there exists a $u_h \in U_h$ such that $f(x_h,u_h) + v  \in \Disc_{{G}_h^{r}}(K_h) + rB_\infty$. The same Lipschitz argument as in the proof of Proposition \ref{prop:PropertiesAlgo}, allows to conclude that for all $\hat{x} \in x_h + rB_\infty$, there exists a $u_h \in U_h$ such that $f(\hat{x},u_h) \in \Disc_{{G}^r_h}(K_h)+rB_\infty$
\end{IEEEproof} 

We conclude our discussion by pointing out that if instead of the discriminating kernel the leadership kernel algorithm is used, then the inner approximation is directly guaranteed. Because by virtue of Assumption \ref{as:grid}, we know that for any robust control input $\bigcup_{I_{u,v}(x_h)} \tilde{V}(u_h,v_h) = V$.

\section{Stationary velocity grid} \label{app:N_m}
The stationary velocity manifold is gridded in $v_x$ and $\delta$ coordinates. In the $v_x$ direction the space is uniformly gridded between a lower and an upper velocity, with a fixed spacing, see Table \ref{tab:gridNM}. For every $v_x$ grid point, $\delta$ is gridded within the normal driving region (see \cite{Liniger_2014}) with a fixed number of grid points, see Table \ref{tab:gridNM}. Additionally, in all cases the same 12 stationary velocity points corresponding to drifting are included.
\begin{table}[h]
\caption{Used gridding variables, for corresponding $N_m$}
\label{tab:gridNM}
\centering
\ra{1.3}
\begin{tabular}{@{}c c c c c @{}}\toprule
$\#\,$modes 	&$v_x$ range & $v_x$ spacing & $\delta$ grid points & drifting modes \\
 \midrule
89   & $[0.5,3.5]$ & 0.25 & 5 & 12 \\ 
99   & $[0.6,3.4]$ & 0.2 & 5 & 12 \\ 
115   & $[0.5,3.5]$ & 0.25 & 7 & 12 \\ 
129   & $ [0.6,3.4]$ & 0.2 & 7 & 12 \\ 
\bottomrule
\end{tabular}
\end{table} 
\end{appendices}

\bibliographystyle{ieeetr}
\bibliography{SpaceDiscRobustViab}

\end{document}